\documentclass[pdftex,twocolumn,epjc3]{svjour3}          

\RequirePackage[T1]{fontenc}

\smartqed  

\RequirePackage{graphicx}
\RequirePackage{mathptmx}      
\RequirePackage{flushend}
\RequirePackage[numbers,sort&compress]{natbib}
\RequirePackage[colorlinks,citecolor=blue,urlcolor=blue,linkcolor=blue]{hyperref}

\usepackage{float}
\usepackage{subcaption}
\usepackage{amsmath,amssymb}

\journalname{Eur. Phys. J. C}

\begin{document}

\title{Greybody factors for massive scalar field emitted from black holes in dRGT massive gravity}


\author{Petarpa Boonserm\thanksref{addr1,e1}
        \and
        Sattha Phalungsongsathit\thanksref{addr2,e2} 
        \and
        Kunlapat Sansuk\thanksref{addr1,e3}
        \and
        Pitayuth Wongjun\thanksref{addr3,e4}
}

\thankstext{e1}{e-mail: Petarpa.Boonserm@gmail.com}
\thankstext{e2}{e-mail: sattha.phalungsongsathit@gmail.com}
\thankstext{e3}{e-mail: kunlapat11032540@gmail.com}
\thankstext{e4}{e-mail: pitbaa@gmail.com}

\institute{Department of Mathematics and Computer Science, Faculty of Science,
Chulalongkorn University, Bangkok 10330, Thailand. \label{addr1}
           \and
           Physics Program, Faculty of Science, Chandrakasem Rajabhat University, Bangkok 10900, Thailand. \label{addr2}
           \and
           The Institute for Fundamental Study “The Tah Poe Academia Institute", 2nd Floor, Mahathamracha Building A, Naresuan University, Mueang Phitsanulok 65000, Thailand. \label{addr3}
}

\date{Received: date / Accepted: date}

\maketitle

\begin{abstract}
Greybody factors are transmission probabilities of the Hawking radiation, which are emitted from black holes and can be obtained from the gravitational potential of black holes. The de Rham, Gabadadze, and Tolly (dRGT) massive gravity is one of the gravity theories that modified general relativity. In this paper, we investigate the greybody factor from the massive scalar field in both the asymptotically dS and the AdS spacetime using the WKB and the rigorous bound methods. We found that the greybody factor depends on the shape of the potential as found in quantum mechanics. The higher the potential barrier, the lower the amount of the grebody factor. Interestingly, for the low multipole case, we found that there exists a critical mass which provides the maximum bound of the greybody factor.  This is a crucial feature of the massive scalar field on the greybody factor from the black holes in both the asymptotically dS and the AdS spacetime.
\end{abstract}

\section{Introduction}
\label{intro}
A classical black hole can only absorb particles but not emit them. However, due to quantum effects, a black hole can create and emit particles from its event horizon, which can be seen as a thermal radiation called the Hawking radiation \cite{Hawking}. This radiation encounters the gravitational potential generated by the black hole itself. This results in the reflection and transmission of the Hawking radiation. Therefore, the actual spectrum observed by an asymptotic observer is different from a blackbody spectrum. A black hole greybody factor is a quantity that describes the deviation of the Hawking radiation from a pure blackbody radiation.

There are various methods to compute the greybody factor. One of them is the general rigorous bound for reflection and transmission coefficients for one-dimensional potential scattering \cite{Visser,P2,Shabat}. This formulation can be applied to a black hole greybody factor \cite{schbh,RN,nonbh,dirty,spin1,KN,Myers,dilatonic,TN,spin0,Myers2,drgt,Myers3,Myers4,bs,perid,fermion}.

Almost all studies on greybody factor are conducted for massless scalar fields. The Massive scalar field is expected to have a different behavior from the massless one in several astrophysical phenomena. For example, it was shown that the massive scalar field has less damping rate than the massless one in a Schwarzschild black hole background \cite{Konoplya}. The mass of the scalar field can cause the instability of the scalar field itself \cite{Furuhashi,Vieira} and the superradiant instability \cite{Bekenstein}. Unlike the massless scalar field, the quasi-normal modes from the massive scalar field can disappear, leaving undamping modes, which are called the quasi-resonance modes when the mass of the scalar field is more than some critical values \cite{Ohashi}.

Furthermore, the massive scalar fields play an important role in elementary particle physics. For example, in the Kaluza-Klein models, the behavior of a massless scalar field in the Fourier modes is similar to the massive one. In black hole spacetimes, the dynamic of the massive scalar fields become important \cite{Koyama}. In this work, we investigate the effects of the massive scalar fields on greybody factors.

One of the alternative theories of gravity is the massive gravity of which the main feature is to give mass to a graviton \cite{Deser,Bergshoeff,Ogievetsky,Mukohyama,Curtright,Alshal}. The ghost-free massive gravity in the four-dimensional spacetime and higher was achieved by de Rham, Gabadadze, and Tolley (dRGT) \cite{deRham,deRham2}. The black hole solutions and their thermodynamics in dRGT massive gravity were found in \cite{Ghosh}. Moreover, the greybody factors for black holes in the asymptotically de Sitter (dS) spacetime in dRGT massive gravity were studied in \cite{drgtbh}.

In this paper, we generalize \cite{drgtbh} to study the greybody factors for the massive scalar fields emitted from black holes in dRGT massive gravity. The dRGT massive gravity is outlined in section \ref{drgt}.  For the dRGT black hole, it is possible to have two horizons for the asymptotically dS spactime and up to three horizons for the asymptotically AdS spactime. We parameterize the existence of the horizon by two model parameters $ 0< \beta_m <1, c_2 <0 $ for  the asymptotically dS black hole and $ 1< \beta_m <\sqrt{4/3}, c_2 >0 $ for the asymptotically AdS spacetime. The potentials of the dRGT black holes are derived in section \ref{poten}. We found that in addition to such two parameters, the potential depends on the scalar field mass. The behaviour of the potential is characterized by using these three parameters in this section. The greybody factors of the dRGT black holes are evaluated for both the rigorous bound and the WKB methods in section \ref{grey}. We found that the WKB method will be efficient for high potential or high multipole, while the results from the rigorous bound will still be valid for all range of parameters. However, the results from the rigorous bound can be approximated to ones from WKB only for the low frequency case. By analyzing the behaviour of the greybody factor, we found that it depends on the shape of the potential as found in quantum mechanics, specifically, the higher the potential barrier, the lesser the amount of the grebody factor. Moreover, for the low multipole case, we found that there exists a critical mass which provides the maximum bound of the greybody factor. Finally, the conclusions and discussions are given in section \ref{con}.

\section{dRGT black hole}\label{drgt}
One of the interesting ways to modify the theory of gravitation  is by adding the graviton mass terms to general relativity which is usually known as massive gravity theory. There are several ways to provide the mass terms. However, most of them are not a good candidate since they encounter the ghost instabilities. Recently, the ghost-free massive gravity theory was proposed by de Rham, Gabadaze and Tolley called the de Rham-Gabadaz-Tolley (dRGT) massive gravity theory, which is represented by the action
\begin{eqnarray}
S &=& \int d^{4}x\sqrt{-g}\left[\frac{M_{P}^{2}}{2}R[g] + m_{g}^{2}(\mathcal{L}_{2}[g, f] + \alpha_{3}\mathcal{L}_{3}[g, f]\right.\nonumber\\
  &&  \left. + \alpha_{4}\mathcal{L}_{4}[g, f])\right],
\end{eqnarray}
where $\alpha_3$ and $\alpha_4$ are model parameters. Note that $R$ is a Ricci scalar, and $m_{g}$ is a graviton mass. The interaction terms denoted by $\mathcal{L}_{i}$s are constructed with the fiducial metric $f_{\mu\nu}$ as
\begin{eqnarray}
\mathcal{L}_{2}[g, f] &=& \frac{1}{2}\left([\mathcal{K}]^{2} - [\mathcal{K}^{2}]\right)\\
\mathcal{L}_{3}[g, f] &=& \frac{1}{3!}\left([\mathcal{K}]^{3} - 3[\mathcal{K}][\mathcal{K}^{2}] + 2[\mathcal{K}^{3}]\right)\\
\mathcal{L}_{4}[g, f] &=& \frac{1}{4!}\left([\mathcal{K}]^{4} - 6[\mathcal{K}]^{2}[\mathcal{K}^{2}] + 3[\mathcal{K}^{2}][\mathcal{K}]^{2}\right.\nonumber\\
                      &&  \left. + 8[\mathcal{K}][\mathcal{K}^{3}] - 6[\mathcal{K}^{4}]\right),
\end{eqnarray}
where the square bracket stands for the trace of a matrix such as $[\mathcal{K}] = g^{\mu\nu}\mathcal{K}_{\mu\nu}$. The tensor $\mathcal{K}_{\mu\nu}$ can be expressed as
\begin{equation}
\mathcal{K}^{\mu}_{\nu} = \delta^{\mu}_{\nu} - \mathcal{X}^{\mu}_{\nu},
\end{equation}
where $\mathcal{X}^{\mu}_{\nu}$ is defined by
\begin{equation}
\mathcal{X}^{\mu}_{\alpha}\mathcal{X}^{\alpha}_{\nu} = g^{\mu\alpha}f_{\nu\alpha}.
\end{equation}
The form of the fiducial metric does not alter the existence of the ghost by construction. It is convenient to choose the form of the fiducial metric as \cite{Volkov,Zhang,Hendi}
\begin{equation}
f_{\mu\nu} = \textrm{diag}(0, 0, c^{2}, c^{2}\sin^{2}\theta), \label{fmunu}
\end{equation}
where $c$ is a constant. The static and the spherically symmetric black hole solution in the dRGT massive gravity is given by \cite{drgtbh}
\begin{equation}
ds^{2} = -f(\tilde{r})dt^{2} + \frac{dr^{2}}{f(\tilde{r})} + r^{2}d\Omega^{2},\label{ds2}
\end{equation}
where $d\Omega^{2} = d\theta^{2} + \sin^{2}\theta d\phi^{2}$, $\tilde{r} = r/c$,
\begin{equation}
f(\tilde{r}) = 1 - \frac{2\tilde{M}}{\tilde{r}} + \alpha_{g}(c_{2} \tilde{r}^{2} - c_{1}\tilde{r} + c_{0}), \label{fr}
\end{equation}
$\tilde{M} = M/c$, $M$ is the black hole mass, $\alpha_{g} = m_{g}^{2}c^{2}$ and $c_{0}, c_{1}$ and $c_{2}$ are dimensionless model parameters related to the parameters $m_{g}^{2}, \alpha_{3}$ and $\alpha_{4}$. By fixing $\alpha_{g}$ to be a positive constant, the $c_{2}$ characterizes the strength of the cosmological constant where it is negative/positive corresponding to the Schwarzschild de Sitter (SdS)/Schwarzschild Anti de Sitter (SAdS) solutions. The parameters $c_0$ and $c_1$ characterize the structure of the interaction terms, which determine the deviation from SdS (SAdS) solutions. Note that the dRGT solution reduces to SdS/SAdS solution in the limit $c_{0} = c_{1} = 0$. Since the parameter $c$ is full, in this solution, it represents a length scale at which the theory manifests the effect of the gravitational modification for $r\gg c$, while the theory reduces to general relativity for $r\ll c$. Indeed, the parameter $c$ is the Vainshtein radius in terms of the screening mechanism.  In this paper, we focus on the case where the black hole has two horizons for the dRGT solution with asymptotically dS spacetime, and three horizons for the dRGT solution with asymptotically AdS spacetime. In order to characterize the existence of the horizons, it is convenient for us to choose the parameter $c_{1}$ as \cite{drgtbh}
\begin{equation}
c_{1} = 3\sqrt[3]{4c_{2}^{2}}.
\end{equation}
Since the parameters are evaluated differently for the dRGT solution with asymptotically dS and AdS spacetime, we will separate our consideration into two subsections below.

\subsection{dS solutions}
\begin{figure}[H]
\centering
  \includegraphics[width=.8\linewidth]{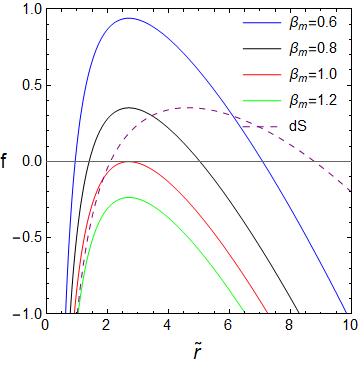}
\caption{\label{frds}The plot of $f_{\textrm{dS}}(\tilde{r})$ with $\tilde{M} = 1$, $c_{2} = -0.01$, $\alpha_{g} = 1$ with various $\beta_m$. The dashed purple line represents $f_{\textrm{SdS}}(\tilde{r})$ ($c_{1} = c_{0} = 0$).}
\end{figure}

Since the dS solution corresponds to $c_{2} < 0$, it is convenient to characterize the existence of the two horizons by 
 redefining the parameter as follows 
\begin{equation}
c_{0} = 3\sqrt{3}\frac{\sqrt[3]{-2c_{2}}}{\beta_{m}} - \frac{1}{\alpha_{g}},
\end{equation}
where the condition for having two horizons can be expressed as
\begin{equation}
0 < \beta_{m} < 1.
\end{equation}
Substituting $c_{1}$ and $c_{0}$ into $f(\tilde{r})$ in equation (\ref{fr}), we obtain
\begin{equation}
f_{\textrm{dS}}(\tilde{r}) = \nonumber
\end{equation}
\begin{equation}
1 - \frac{2\tilde{M}}{\tilde{r}} + \alpha_{g}\left(c_{2} \tilde{r}^{2} - 3\sqrt[3]{4c_{2}^{2}}\tilde{r} + 3\sqrt{3}\frac{\sqrt[3]{-2c_{2}}}{\beta_{m}} - \frac{1}{\alpha_{g}}\right). \label{newfr}
\end{equation}
The horizons can be obtained by solving equation $f_{\textrm{dS}}(\tilde{r})$ = 0. As a result, such two horizons can be written as
\begin{equation}
\tilde{r}_{1} = \frac{2}{\sqrt[3]{-2c_{2}}}\left[\sqrt{X}\cos\left(\frac{1}{3}\textrm{arcsec}(Y)\right) - 1\right]
\end{equation}
and
\begin{equation}
\tilde{r}_{2} = -\frac{2}{\sqrt[3]{-2c_{2}}}\left[\sqrt{X}\cos\left(\frac{1}{3}\textrm{arcsec}(Y) + \frac{\pi}{3}\right) + 1\right],
\end{equation}
where 
\begin{equation}
X = \frac{2\sqrt{3}}{\beta_{m}} + 4 ~~\textrm{and}~~ Y = -\sqrt{\frac{2}{\beta_{m}}}\frac{(2\beta_{m} + \sqrt{3})^{3/2}}{5\beta_{m} + 3\sqrt{3}}.
\end{equation}
From Fig.\ref{frds}, one can see that the parameter $\beta_m$ can be used to characterize the existence to two horizons. Moreover, for this setting of parameters, one can find that two horizons in the dRGT black hole are closer than ones in the SdS black hole. This is one of the important properties of the dRGT black hole, which provides a crucial contribution to the greybody factor in the dRGT black hole, which is different from one in the SdS black hole.

\subsection{AdS solutions} 
\begin{figure}[H]
\centering
  \includegraphics[width=.8\linewidth]{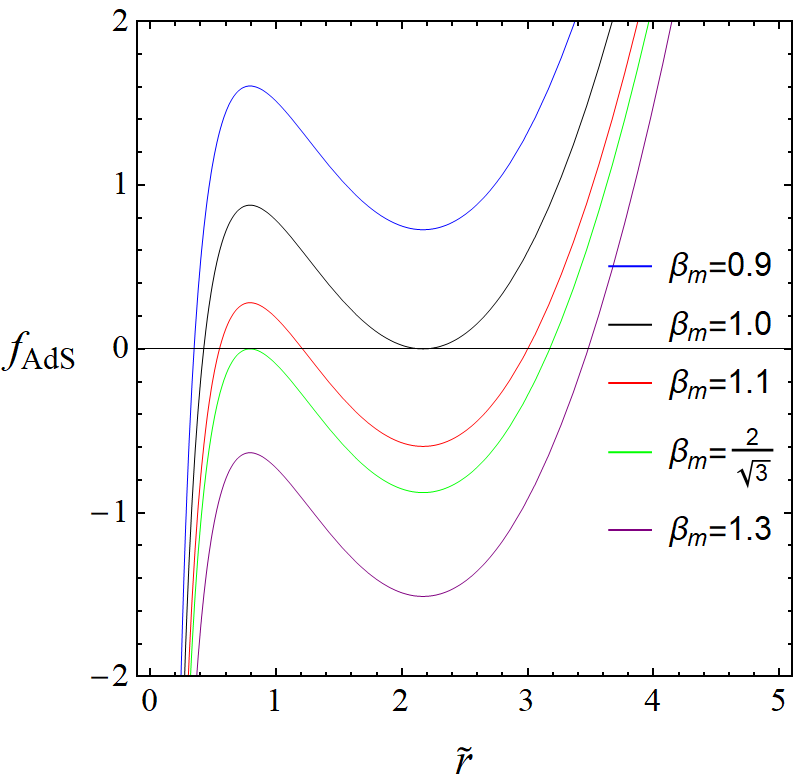}
\caption{\label{frads}The plot of $f(\tilde{r})$ with $\tilde{M} = 1$, $c_{2} = 1$, $\alpha_{g} = 1$ and $\beta_{m} = 0.9$, 1.0, 1.1, $2/\sqrt{3}$ and 1.}
\end{figure}

For the AdS solutions, the parameter $c_2$ is positive, $c_{2} > 0$. By performing in a similar way as done in the dS case, the parameter $c_0$ can be written in terms of $\beta_m$ as
\begin{equation}
c_{0} = 3\sqrt{3}\frac{\sqrt[3]{2c_{2}}}{\beta_{m}} - \frac{1}{\alpha_{g}},
\end{equation}
where the existence of three horizons can be characterized by parameter $\beta_m$ as
\begin{equation}
1 < \beta_{m} < \frac{2}{\sqrt{3}}. \label{cond3hor}
\end{equation}
Substituting $c_{1}$ and $c_{0}$ into $f(\tilde{r})$ in equation (\ref{fr}), we obtain
\begin{equation}
f_{\textrm{AdS}}(\tilde{r}) = \nonumber
\end{equation}
\begin{equation}
1 - \frac{2\tilde{M}}{\tilde{r}} + \alpha_{g}\left(c_{2} \tilde{r}^{2} - 3\sqrt[3]{4c_{2}^{2}}\tilde{r} + 3\sqrt{3}\frac{\sqrt[3]{2c_{2}}}{\beta_{m}} - \frac{1}{\alpha_{g}}\right). \label{newfrads}
\end{equation}
By solving $f_{\textrm{AdS}}(\tilde{r}) = 0$, the three horizons can be expressed as
\begin{equation}
\tilde{r}_{1} = \frac{2}{\sqrt[3]{2c_{2}}}\left[1 - \sqrt{x}\sin\left(\frac{1}{3}\textrm{arcsec}(y) + \frac{\pi}{6}\right)\right],
\end{equation}
\begin{equation}
\tilde{r}_{2} = \frac{2}{\sqrt[3]{2c_{2}}}\left[1 - \sqrt{x}\cos\left(\frac{1}{3}\textrm{arcsec}(y) + \frac{\pi}{3}\right)\right],
\end{equation}
\begin{equation}
\tilde{r}_{3} = \frac{2}{\sqrt[3]{2c_{2}}}\left[1 + \sqrt{x}\cos\left(\frac{1}{3}\textrm{arcsec}(y)\right)\right],
\end{equation}
where
\begin{equation}
x = \frac{4 - 2\sqrt{3}}{\beta_{m}} ~~\textrm{and}~~ y = \frac{\sqrt{6} - 2\sqrt{2}\beta_{m}}{(3\sqrt{3} - 5\beta_{m})\sqrt{-\frac{\beta_{m}}{\sqrt{3} - 2\beta_{m}}}}.
\end{equation}
From Fig. \ref{frads}, one can see that the parameter $\beta_m$ can characterize the existence of three horizons, satisfying the condition in  Eq. (\ref{cond3hor}). For $\beta_{m} = 1$, the second and the third horizons emerge so that in this case, it corresponds to extremal black hole with two horizons. For $\beta_{m} = 2/\sqrt{3}$, the first and the second horizons emerge. In this case, the black hole is also extremal with two horizons. Note that, for the SAdS black hole, there exists only the first horizon (similar to the case for $\beta_m < 1$). Therefore, the existence of three horizons is a crucial property of the dRGT black hole compared to the SAdS black hole. 

Moreover, in order to find the proper form of the potential in the next section, it is worthwhile to find the derivative of the horizon function. As a result, we found that $f'_{\textrm{dS}}(\tilde{r})= f'_{\textrm{AdS}}(\tilde{r})$ and can be written as 
\begin{equation}
f'_{\textrm{dS}}(\tilde{r}) = f'_{\textrm{AdS}}(\tilde{r}) = \frac{2\tilde{M}}{\tilde{r}^{2}} + 2\alpha_{g}c_{2}\tilde{r} - 3\alpha_{g}\sqrt[3]{4c_{2}^{2}}.
\end{equation}
Note that the signs of $c_{2}$ are different; for the dS case, $c_{2}$ is negative, while for the AdS case, $c_{2}$ is positive.

\section{Potential of the dRGT black hole}\label{poten}
When quantum effects are taken into account, a black hole can emit thermal radiation at its event horizon called the Hawking radiation \cite{Hawking}. This radiation could be any field such as the scalar field or the electromagnetic field. In this work, we are interested in the massive scalar field emitted from a black hole. Its dynamics can be described by the Klein-Gordon equation on a curved spacetime \cite{Fulling}
\begin{equation}
\frac{1}{\sqrt{-g}}\partial_{\mu}(\sqrt{-g}g^{\mu\nu}\partial_{\nu} \Phi) - m^{2}\Phi = 0,
\end{equation}
where $g_{\mu\nu}$ is the metric tensor, $g^{\mu\nu}$ is the inverse of the metric tensor, $g$ is the determinant of the metric tensor and $m$ is the mass of the scalar field.

Since a black hole from equation (\ref{ds2}) has spherical symmetry, the method of the separation of variables can be used. We write $\Phi$ as
\begin{equation}
\Phi(t, r, \theta, \phi) = e^{i\omega t}\frac{\psi(r)}{r}Y_{\ell m}(\theta, \phi),
\end{equation}
where $Y_{\ell m}(\theta, \phi)$ is a spherical harmonic function. It satisfies
\begin{equation}
\frac{1}{\sin\theta}\frac{\partial}{\partial\theta}\left(\sin\theta\frac{\partial Y_{\ell m}(\theta, \phi)}{\partial\theta}\right) + \frac{1}{\sin^{2}\theta}\frac{\partial^{2}Y_{\ell m}(\theta, \phi)}{\partial\phi^{2}} = \nonumber
\end{equation}
\begin{equation}
-\ell(\ell + 1)Y_{\ell m}(\theta, \phi),
\end{equation}
where $\ell$ is a non-negative integer, $m$ is an integer and $\ell \geq |m|$. The radial part is given by
\begin{equation}
\frac{d^{2}\psi(r)}{dr_{*}^{2}} + [\omega^{2} - V(\tilde{r})]\psi(r) = 0, \label{rp}
\end{equation} \label{RE}
where
\begin{equation}
\frac{dr_{*}}{dr} = \frac{1}{|f(\tilde{r})|} \label{dr*}
\end{equation}
and
\begin{equation}
\tilde{V}(\tilde{r}) = f(\tilde{r})\left[\frac{\ell(\ell + 1)}{\tilde{r}^{2}} + \frac{f'(\tilde{r})}{\tilde{r}} + \tilde{m}^{2}\right]. \label{V}
\end{equation}
where $\tilde{V}(\tilde{r}) = c^{2}V(\tilde{r})$ and $\tilde{m} = cm$.

\subsection{Potentials for the dS case}
\begin{figure}[H]
\begin{subfigure}{.45\textwidth}
  \centering
  \includegraphics[width=.8\linewidth]{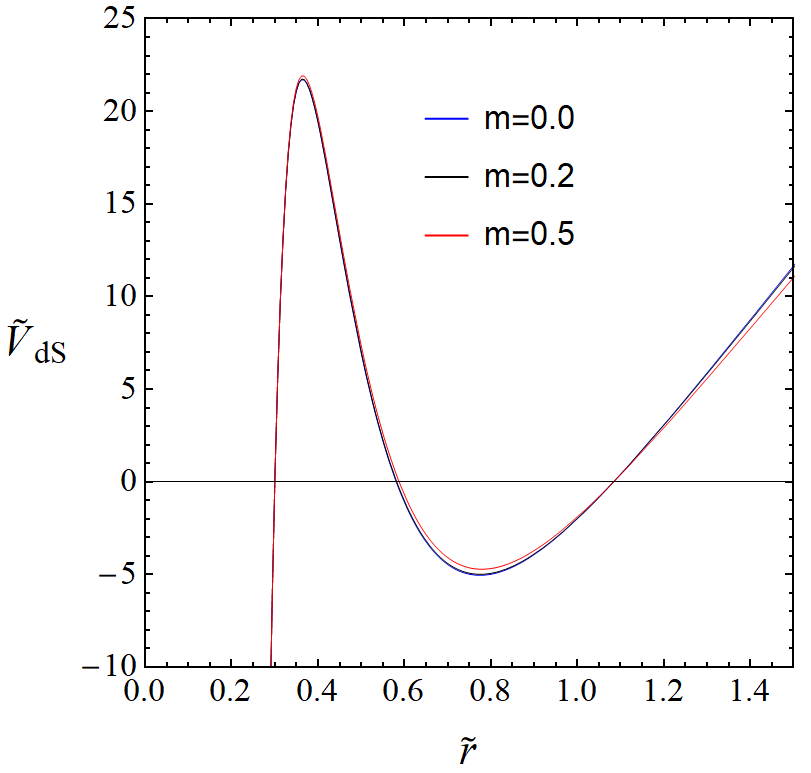}
  \caption{}
\end{subfigure}
\begin{subfigure}{.45\textwidth}
  \centering
  \includegraphics[width=.8\linewidth]{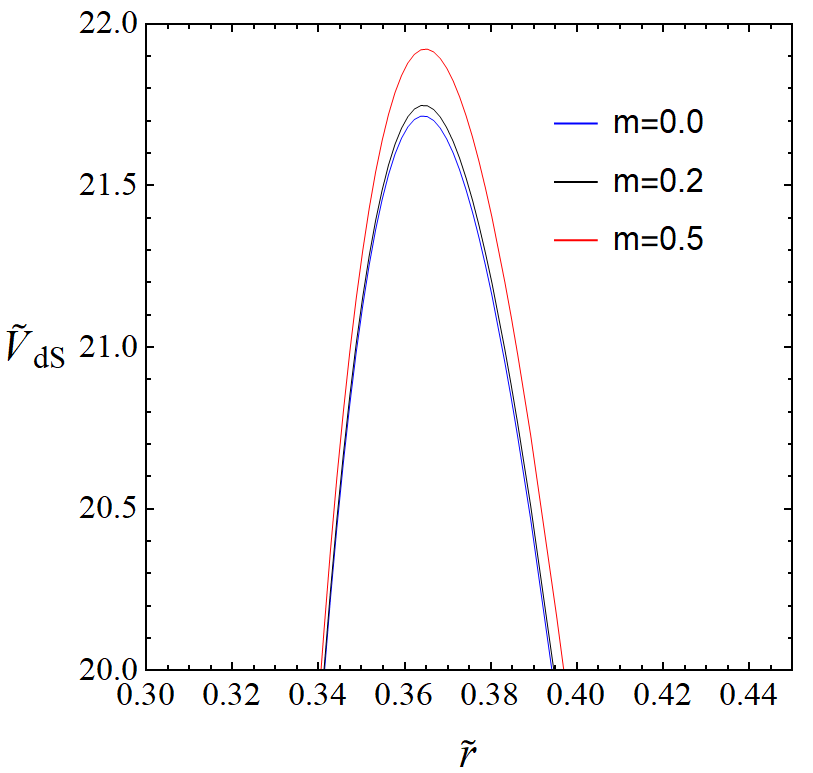}
  \caption{}
\end{subfigure}
\caption{\label{pot}(a) The dRGT potentials for the dS case with $\ell = 0$, $\tilde{M} = 1$, $c_{2} = -1$, $\alpha_{g} = 1$, $\beta_{m} = 0.8$ and $m = 0$, 0.2 and 0.5. (b) The magnification of (a).}
\end{figure}

From equation (\ref{V}), the dS potential is given by
\begin{equation}
\tilde{V}_{\textrm{dS}}(\tilde{r}) = f_{\textrm{dS}}(\tilde{r})\left[\frac{\ell(\ell + 1)}{\tilde{r}^{2}} + \frac{f'_{\textrm{dS}}(\tilde{r})}{\tilde{r}} + \tilde{m}^{2}\right], \label{Vds}
\end{equation}
where $f_{\textrm{dS}}(\tilde{r})$ is given in equation (\ref{newfr}). The potentials for the dS case are plotted for $m = 0$, 0.2 and 0.5, with different $\beta_{m}$ in Figures \ref{pot}. The graph shows that when $m$ increases, the potential also increases. This is apparent from equation (\ref{Vds}) which shows that the potential is an increasing function of $m$ because $f_{\textrm{dS}}(\tilde{r})$ is always positive. Therefore, when $m$ increases, the whole graph is shifted upward as shown in Figures \ref{pot}.

For $\ell = 0$, the dS potential for massless scalar field vanishes at $\tilde{r}_{\textrm{fmax}}$, where $\tilde{r}_{\textrm{fmax}}$ is the location of the maximum of $f_{\textrm{dS}}(\tilde{r})$ ($f'_{\textrm{dS}}(\tilde{r}_{\textrm{fmax}}) = 0$). However, for the massive scalar field, it does not vanish at this point, but is still positive and vanishes at $\tilde{r} > \tilde{r}_{\textrm{fmax}}$. We can see this from Figure \ref{pot} that the $\tilde{r}$ intercept of the red curve ($m$ = 0.5) is on the right of that of the blue curve ($m = 0$). Thus, the mass of the scalar field has effects on the area under the curve of the potential. The scalar mass increases the area which is above the $\tilde{r}$ axis, and decrease the area which is below the $\tilde{r}$ axis.

\begin{figure}[H]
\begin{subfigure}{.45\textwidth}
  \centering
  \includegraphics[width=.8\linewidth]{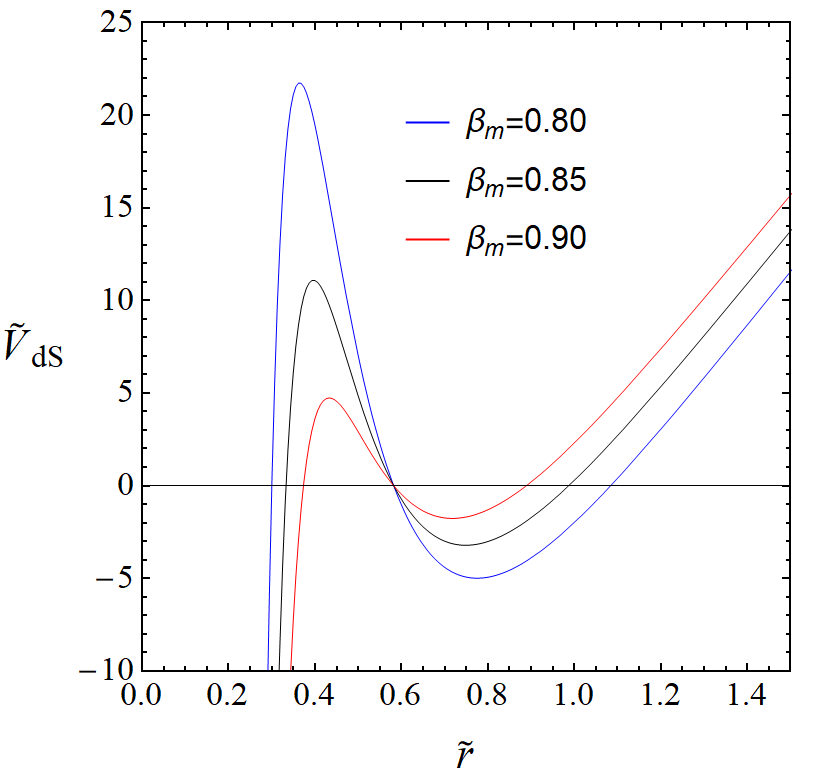}
  \caption{}
\end{subfigure}
\begin{subfigure}{.45\textwidth}
  \centering
  \includegraphics[width=.8\linewidth]{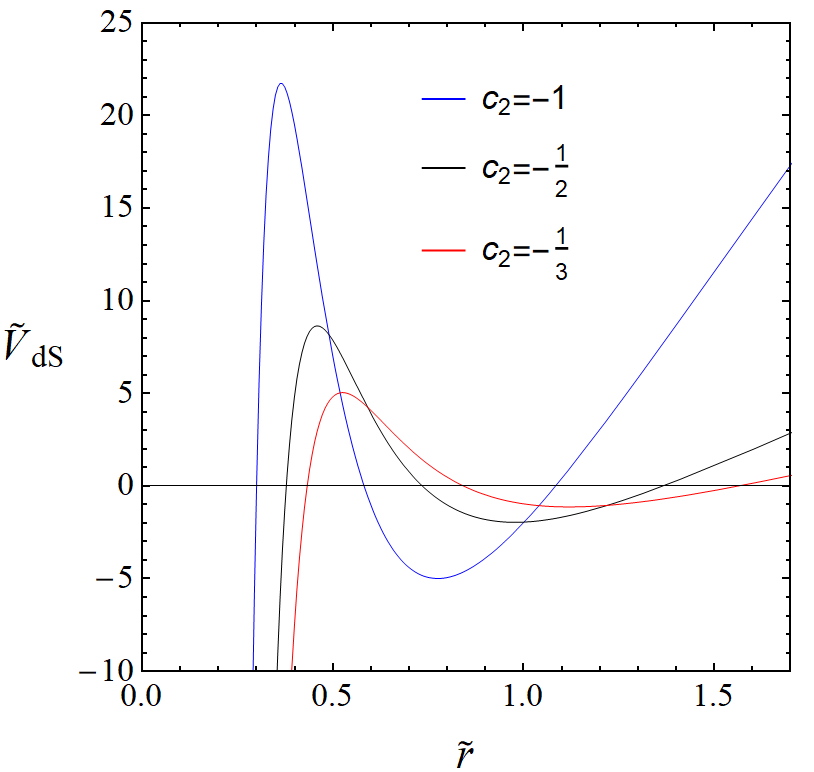}
  \caption{}
\end{subfigure}
\caption{\label{pot2}The dRGT potentials for the dS case with $\ell = 0$, $\tilde{M} = 1$, $\alpha_{g} = 1$, and $m = 0.2$. (a) $c_{2} = -1$ and $\beta_{m} = 0.8$, 0.85 and 0.9. (b) $\beta_{m} = 0.8$ and $c_{2} = -1$, -1/2 and -1/3.}
\end{figure}

The dS potentials are plotted with different $\beta_{m}$ and $c_{2}$ in Figures \ref{pot2}(a) and \ref{pot2}(b), respectively. From Figure \ref{pot2}(a), we see that when $\beta_{m}$ increases, the local maximum of the potential decreases, while the local minimum increases. This can be seen by finding the derivative of the potential in equation (\ref{Vds}) with respect to $\beta_{m}$
\begin{equation}
\frac{d\tilde{V}_{\textrm{dS}}}{d\beta_{m}} = -3\sqrt{3}\alpha_{g}\frac{\sqrt[3]{-2c_{2}}}{\beta_{m}^{2}}\left[\frac{\ell(\ell + 1)}{\tilde{r}^{2}} + \frac{f'_{\textrm{dS}}(\tilde{r})}{\tilde{r}} + \tilde{m}^{2}\right].
\end{equation}
If the sum of all terms in the square bracket is positive, we obtain $d\tilde{V}_{\textrm{dS}}/d\beta_{m} < 0$, and the potential decreases with increasing $\beta_{m}$. On the other hand, if the sum of all terms in the square bracket is negative, the potential increases with increasing $\beta_{m}$. Since $f_{\textrm{dS}}(\tilde{r})$ is always positive, from equation (\ref{Vds}), the potential has the same sign as the square bracket term. Therefore, when the potential is positive, corresponding to the positive square bracket term, it decreases with increasing $\beta_{m}$. We can see from Figure \ref{pot2}(a) that the red curve ($\beta_{m} = 0.9$) is below the blue curve ($\beta_{m} = 0.8$) in the region where the potential is positive. Conversely, when the potential is negative, corresponding to the negative square bracket term, it increases with increasing $\beta_{m}$. Again, we can see from Figure \ref{pot2}(a) that the red curve ($\beta_{m} = 0.9$) is above the blue curve ($\beta_{m} = 0.8$) in the region where the potential is negative. Moreover, when $\beta_{m}$ increases, the width of the potential decreases.

For Figure \ref{pot2}(b), we see that when $c_{2}$ decreases, the local maximum of the potential increases, while the local minimum decreases. Since $c_{2}$ represents the negative of the cosmological constant, the decrease in $c_{2}$ for the dS case corresponds to the increase in the magnitude of the positive cosmological constant. Therefore, the graph shows that the local maximum of the potential increases when the magnitude of the cosmological constant increases.

\begin{figure}[H]
\centering
\includegraphics[width=.48\linewidth]{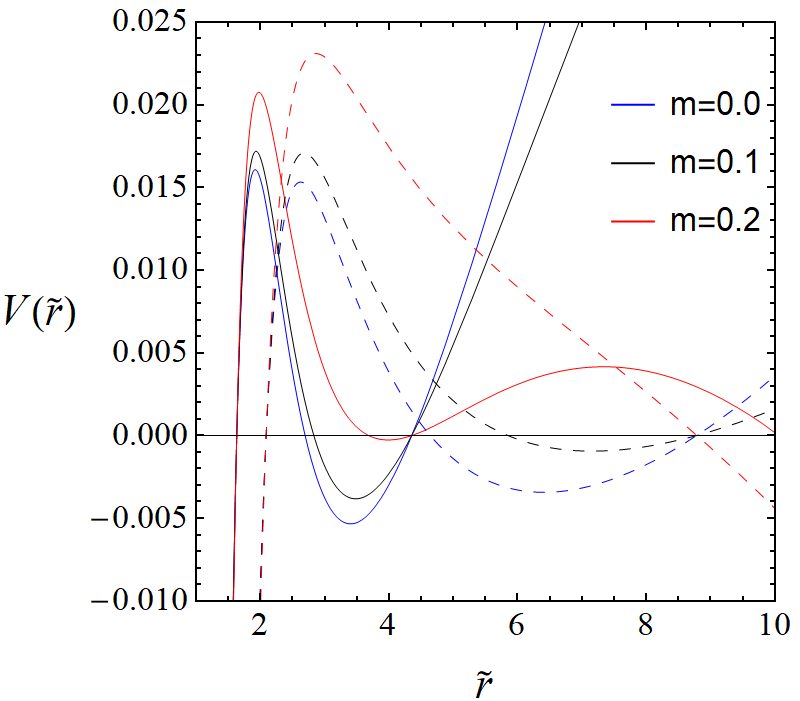}\quad
\includegraphics[width=.48\linewidth]{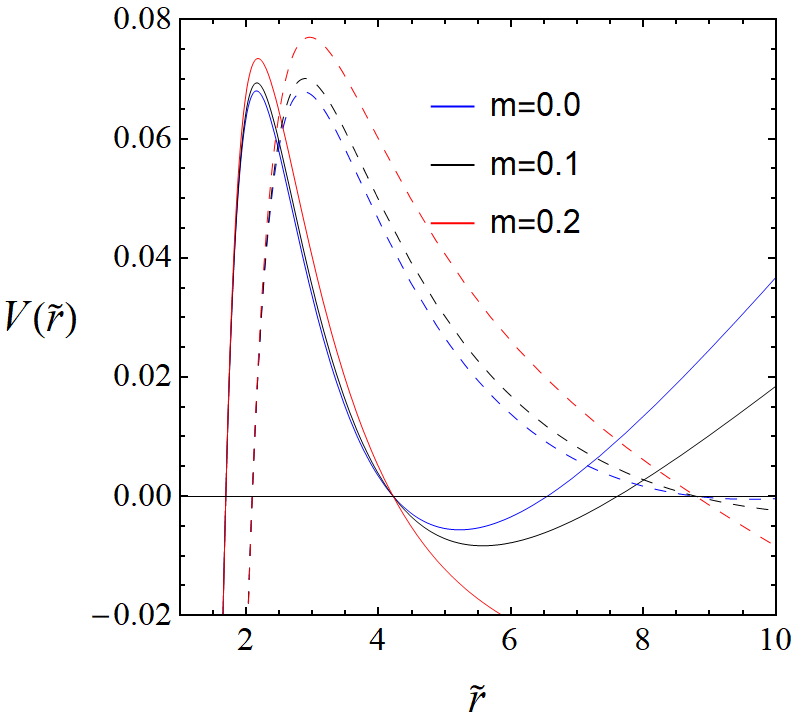}
\caption{\label{vdsdrgt}The potentials with ($\ell = 0$ left panel and  $\ell = 1$ right panel), $\tilde{M} = 1$, $c_{2} = -0.01$, $\alpha_{g} = 1$. The solid lines represent the potential for an asymptotically dS black hole in dRGT massive gravity ($\beta_{m} = 0.875$ for left panel and $\beta_{m} = 0.89$ for right panel). The dashed lines represent the potential for an SdS black hole ($c_{1} = c_{0} = 0$).}
\end{figure}

When $c_{1} = c_{0} = 0$, the asymptotically dS spacetime in the dRGT massive gravity reduces to the Schwarzs-child de Sitter (SdS) spacetime. The function $f(\tilde{r})$ of the SdS spacetime can be obtained by letting $c_{1} = c_{0} = 0$ in equation (\ref{fr})
\begin{equation}
f_{\textrm{SdS}}(\tilde{r}) = 1 - \frac{2\tilde{M}}{\tilde{r}} + \alpha_{g}c_{2} \tilde{r}^{2}
\end{equation}
and $c_{2}$ be negative. Then, from equation (\ref{V}), the SdS potential is given by
\begin{equation}
\tilde{V}_{\textrm{SdS}}(\tilde{r}) = f_{\textrm{SdS}}(\tilde{r})\left[\frac{\ell(\ell + 1)}{\tilde{r}^{2}} + \frac{f'_{\textrm{SdS}}(\tilde{r})}{\tilde{r}} + \tilde{m}^{2}\right]. \label{vsds}
\end{equation}
The potentials for an SdS black hole and an asymptotically dS black hole in the dRGT massive gravity are plotted in Figure \ref{vdsdrgt}. We see that the potential for the dS black hole in the dRGT massive gravity (the solid curve) is shifted left from the potential for the SdS black hole (the dashed curve). The massive gravity shrinks the horizons of a black hole, which is apparent from Figure \ref{frds}. Notice that when $m$ increases, the absolute area of the dS potential for $\ell = 0$ in the negative region decreases. This means that if $m$ continues to increase to some value, called $m_{\min}$, the dS potential becomes positive on the interval $\tilde{r}_{1} < \tilde{r} < \tilde{r}_{2}$. To find $m_{\min}$, we rewrite equation (\ref{Vds}) for $\ell = 0$ as
\begin{equation}
\tilde{V}_{\textrm{dS}, \ell = 0}(\tilde{r}) = \nonumber
\end{equation}
\begin{equation}
f_{\textrm{dS}}(\tilde{r})\left(\frac{2\tilde{M}}{\tilde{r}^{3}} + 2\alpha_{g}c_{2} - \frac{3\alpha_{g}\sqrt[3]{4c_{2}^{2}}}{\tilde{r}} + \tilde{m}^{2}\right).
\end{equation}
It is positive for $\tilde{r}_{1} < \tilde{r} < \tilde{r}_{2}$ if
\begin{equation}
\tilde{m}^{2} > -\frac{2\tilde{M}}{\tilde{r}^{3}} - 2\alpha_{g}c_{2} + \frac{3\alpha_{g}\sqrt[3]{4c_{2}^{2}}}{\tilde{r}}.
\end{equation}
We define $\tilde{m}_{\min}$ as
\begin{equation}
\tilde{m}_{\min}^{2} = \max\left[-\frac{2\tilde{M}}{\tilde{r}^{3}} - 2\alpha_{g}c_{2} + \frac{3\alpha_{g}\sqrt[3]{4c_{2}^{2}}}{\tilde{r}}\right].
\end{equation}
where $\tilde{r}_{1} < \tilde{r} < \tilde{r}_{2}$. We find that the maximum is at $\tilde{r} = \tilde{r}_{2}$. Therefore,
\begin{equation}
\tilde{m}_{\min}^{2} = -\frac{2\tilde{M}}{\tilde{r}_{2}^{3}} - 2\alpha_{g}c_{2} + \frac{3\alpha_{g}\sqrt[3]{4c_{2}^{2}}}{\tilde{r}_{2}}. \label{mmin2}
\end{equation}
For $\tilde{M} = 1$, $c_{2} = -0.01$ and $\alpha_{g} = 1$, we obtain
\begin{equation}
\tilde{m}_{\min}^{2} = 0.02 + \frac{0.22}{\tilde{r}_{2}} - \frac{2}{\tilde{r}_{2}^{3}}.
\end{equation}
For $\beta_{m} = 0.875$ and $c_{2} = -0.01$, we obtain $\tilde{r}_{2} = 4.357$. Therefore,
\begin{equation}
\tilde{m}_{\min} = 0.215.
\end{equation}
Moreover, when $\tilde{m} = 0.2$, an SdS potential for $\ell = 0$ is positive on the interval $\tilde{r}_{1} < \tilde{r} < \tilde{r}_{2}$. In the case of the SdS black hole, $\tilde{m}_{\min}$ in equation (\ref{mmin2}) reduces to
\begin{equation}
\tilde{m}_{\min}^{2} = -\frac{2\tilde{M}}{\tilde{r}_{2}^{3}} - 2\alpha_{g}c_{2}.
\end{equation}
For $c_{2} = -0.01$ and $\beta_{m} = 0.875$, which give $\tilde{r}_{2} = 8.79$, and $\tilde{M} = 1$ and $\alpha_{g} = 1$, we obtain
\begin{equation}
\tilde{m}_{\min} = 0.13.
\end{equation}
This is why the SdS potential for $\ell = 0$ in Figure \ref{vdsdrgt} is positive for $\tilde{r}_{1} < \tilde{r} < \tilde{r}_{2}$ at $\tilde{m} = 0.2$.

Furthermore, for $\ell = 1$, the potentials are positive for $\tilde{r}_{1} < \tilde{r} < \tilde{r}_{2}$. To see this, we rewrite equation (\ref{Vds}) as
\begin{equation}
\tilde{V}_{\textrm{dS}}(\tilde{r}) = \nonumber
\end{equation}
\begin{equation}
f_{\textrm{dS}}(\tilde{r})\left[\frac{\ell(\ell + 1)}{\tilde{r}^{2}} + \frac{2\tilde{M}}{\tilde{r}^{3}} + 2\alpha_{g}c_{2} - \frac{3\alpha_{g}\sqrt[3]{4c_{2}^{2}}}{\tilde{r}} + \tilde{m}^{2}\right].
\end{equation}
It is positive for $\tilde{r}_{1} < \tilde{r} < \tilde{r}_{2}$ if
\begin{equation}
\ell(\ell + 1) > -\frac{2\tilde{M}}{\tilde{r}} - 2\alpha_{g}c_{2}\tilde{r}^{2} + 3\alpha_{g}\sqrt[3]{4c_{2}^{2}}\tilde{r} - \tilde{m}^{2}\tilde{r}^{2}.
\end{equation}
From this expression, the left hand side is in the order of unity and the right hand side is typically less than unity by the requirement of the existence of the horizons. For example, for a typical scale of the existence of the horizons, $c_{2} \sim -0.01$ and then  $\tilde{r} \gtrsim 1$. For the usual scale of the scalar mass $\tilde{m}^2 \sim 0.01 $, the second term and the forth term are in the same order, which provide a tiny contribution. The third term is in the order of $0.1 \tilde{r}$ which contributes to the same order as the first term in the opposite sign. Therefore, the amount of contribution of the right hand side is usually smaller than unity. For example, by setting  $\tilde{M}= \alpha_g =1$, $c_{2} = -0.01$, $\beta_m = 0.875$ and  $\tilde{r} \sim \tilde{r}_2 \sim  4.36$, we obtain $\ell(\ell + 1) > 0.0034$ with $m^2 =0.05$. As a result, one can see that it is difficult to obtain the negative part of the potential for the case of $\ell \geq 1$. 

\subsection{Potentials of the AdS case}
From equation (\ref{V}), the AdS potential is given by
\begin{equation}
\tilde{V}_{\textrm{AdS}}(\tilde{r}) = f_{\textrm{AdS}}(\tilde{r})\left[\frac{\ell(\ell + 1)}{\tilde{r}^{2}} + \frac{f'_{\textrm{AdS}}(\tilde{r})}{\tilde{r}} + \tilde{m}^{2}\right], \label{Vads}
\end{equation}
where $f_{\textrm{AdS}}(\tilde{r})$ is given in equation (\ref{newfrads}). In the AdS case, $f_{\textrm{AdS}}(\tilde{r})$ has three horizons. Thus, from equation (\ref{Vads}), the potential is zero at five points of $\tilde{r}$.

\begin{figure}[H]
  \centering
  \includegraphics[width=.8\linewidth]{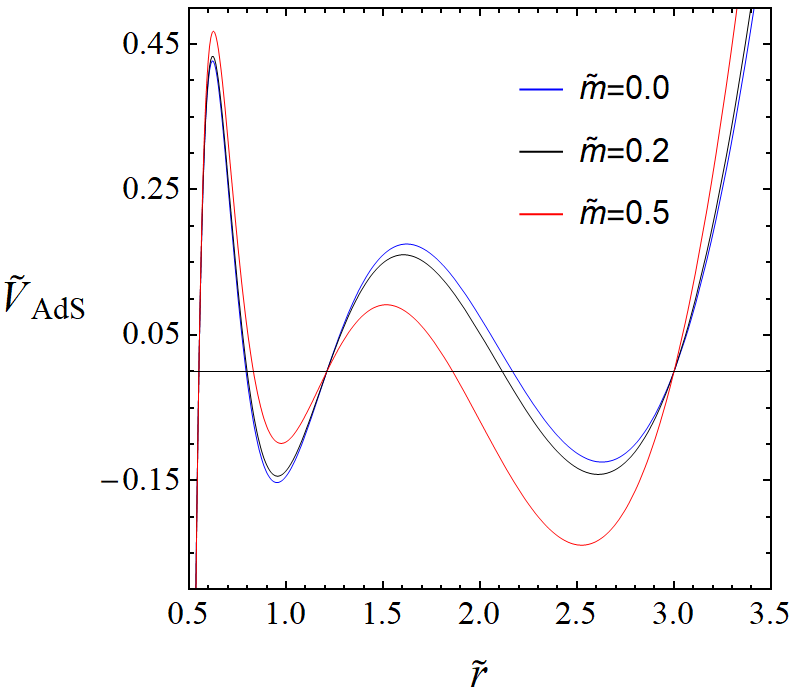}
\caption{\label{pot3}The AdS potentials with $\ell = 0$, $\tilde{M} = 1$, $c_{2} = 1$, $\alpha_{g} = 1$, $\beta_{m} = 1.1$ and $\tilde{m} = 0$, 0.2 and 0.5.}
\end{figure}

The AdS potentials are plotted in Figure \ref{pot3}. Consider the effect of $m$. Different from the dS case, the AdS potential is an increasing function in $m$ when $f_{\textrm{AdS}}(\tilde{r}) > 0$ and a decreasing function in $m$ when $f_{\textrm{AdS}}(\tilde{r}) < 0$. In the range $\tilde{r}_{1} < \tilde{r} < \tilde{r}_{2}$, where $f_{\textrm{AdS}}(\tilde{r})$ is positive, the graph is shifted upward with increasing $m$. On the other hand, in the range $\tilde{r}_{2} < \tilde{r} < \tilde{r}_{3}$, where $f_{\textrm{AdS}}(\tilde{r})$ is negative, the graph is shifted downward with increasing $m$. These are shown in Figure \ref{pot3} where we can see the red curve ($m = 0.5$) is above the blue curve ($m = 0$) in the range $\tilde{r}_{1} < \tilde{r} < \tilde{r}_{2}$, while the red curve is below the blue curve in the range $\tilde{r}_{2} < \tilde{r} < \tilde{r}_{3}$. The mass of scalar field can increase or decrease the potential depending on the sign of $f_{\textrm{AdS}}(\tilde{r})$.

\begin{figure}[H]
\begin{subfigure}{.45\textwidth}
  \centering
  \includegraphics[width=.8\linewidth]{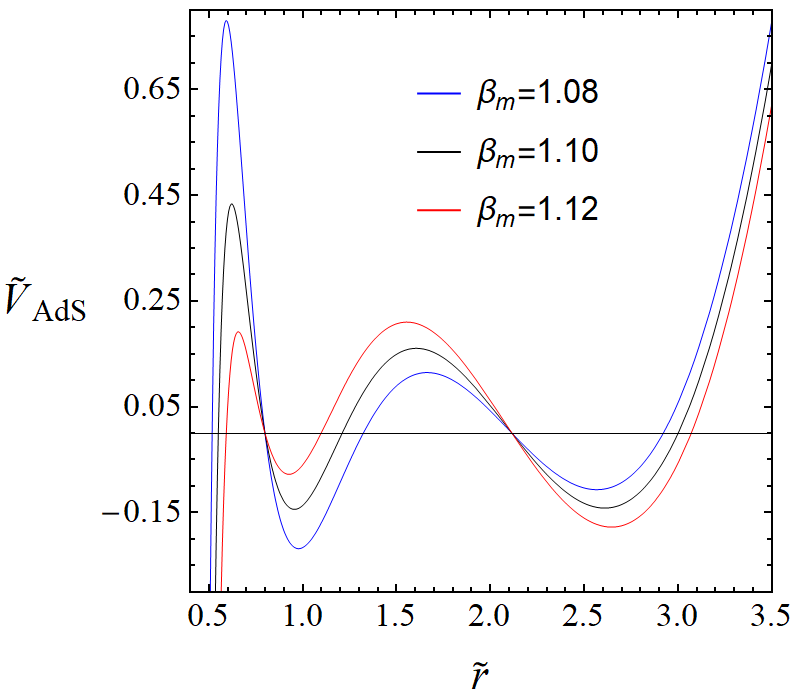}
  \caption{}
\end{subfigure}
\begin{subfigure}{.45\textwidth}
  \centering
  \includegraphics[width=.8\linewidth]{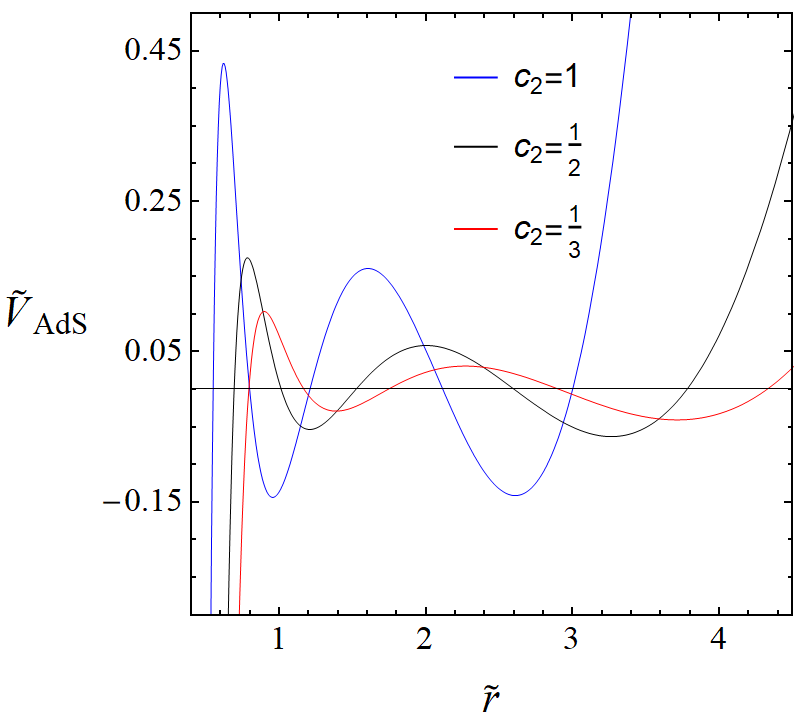}
  \caption{}
\end{subfigure}
\caption{\label{pot4}The AdS potentials with $\ell = 0$, $\tilde{M} = 1$, $\alpha_{g} = 1$, and $\tilde{m} = 0.2$. (a) $c_{2} = 1$ and $\beta_{m} = 1.08$, 1.10 and 1.12. (b) $\beta_{m} = 1.1$ and $c_{2} = 1$, 1/2 and 1/3.}
\end{figure}

The AdS potentials are plotted with different $\beta_{m}$ and $c_{2}$ in Figures \ref{pot4}(a) and \ref{pot4}(b), respectively. We can analyze the effects of $\beta_{m}$ on the potential by finding the derivative of the potential in equation (\ref{Vads}) with respect to $\beta_{m}$
\begin{equation}
\frac{d\tilde{V}_{\textrm{AdS}}}{d\beta_{m}} = -3\sqrt{3}\alpha_{g}\frac{\sqrt[3]{2c_{2}}}{\beta_{m}^{2}}\left[\frac{\ell(\ell + 1)}{\tilde{r}^{2}} + \frac{f'_{\textrm{AdS}}(\tilde{r})}{\tilde{r}} + \tilde{m}^{2}\right].
\end{equation}
Similar to the dS case, if the square bracket term is positive, the potential decreases with increasing $\beta_{m}$. On the other hand, if the square bracket term is negative, the potential increases with increasing $\beta_{m}$. In the range $\tilde{r}_{1} < \tilde{r} < \tilde{r}_{\textrm{max}1}$, the square bracket term is positive. Therefore, the potential in this range decreases with increasing $\beta_{m}$. We can see from Figure \ref{pot4}(a) that the red curve ($\beta_{m} = 1.12$) is below the blue curve ($\beta_{m} = 1.08$). In the range $\tilde{r}_{\textrm{max}1} < \tilde{r} < \tilde{r}_{2}$, the square bracket term is negative. Thus, the potential in this range increases with increasing $\beta_{m}$. We can see that the red curve is above the blue curve. In the range $\tilde{r}_{2} < \tilde{r} < \tilde{r}_{\textrm{max}2}$, the square bracket term is again negative. Then, the potential in this range increases with increasing $\beta_{m}$. We can see that the red curve is above the blue curve. In the range $\tilde{r}_{\textrm{max}2} < \tilde{r} < \tilde{r}_{3}$, the square bracket term is positive and the potential in this range decreases with increasing $\beta_{m}$. We can see that the red curve is below the blue curve.

For Figure \ref{pot4}(b), we see that when $c_{2}$ increases, the relative maximum of the potential also increases. The increase in $c_{2}$ for the AdS case corresponds to the increase in the magnitude of the negative cosmological constant. Therefore, the graph shows that the local maximum of the potential increases when the magnitude of the cosmological constant increases.

\begin{figure}[H]
  \centering
  \includegraphics[width=.8\linewidth]{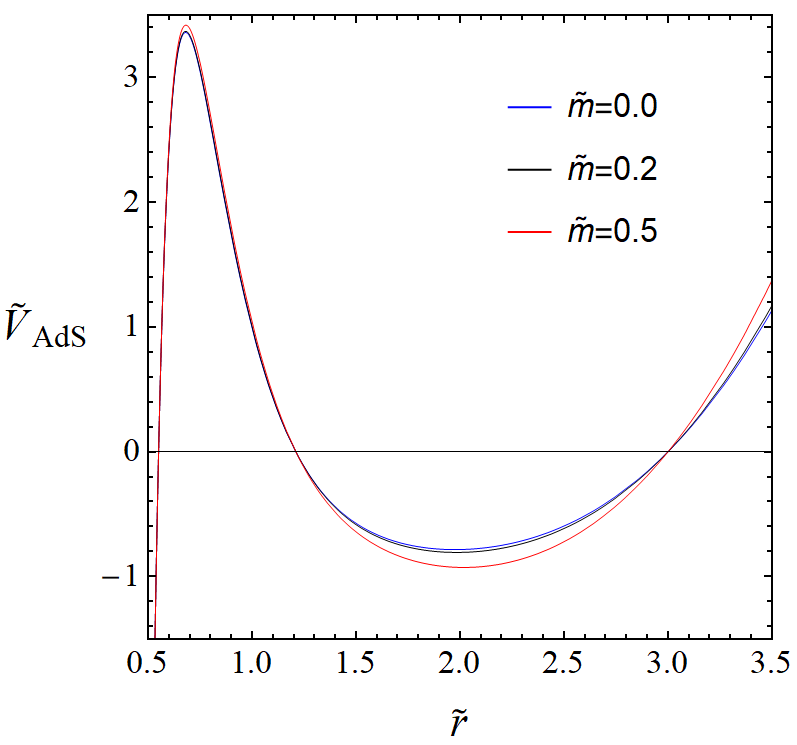}
\caption{\label{potadsl12}The AdS potentials with $c_{2} = 1$, $\beta_{m} = 1.1$, $\tilde{M} = 1$, $\alpha_{g} = 1$, $\tilde{m} = 0$, 0.2 and 0.5 and $\ell = 2$.}
\end{figure}

The AdS potentials for $\ell = 2$ are plotted in Figure \ref{potadsl12}. We see that the potentials are positive on $\tilde{r}_{1} < \tilde{r} < \tilde{r}_{2}$ and negative on $\tilde{r}_{2} < \tilde{r} < \tilde{r}_{3}$.

\section{Greybody factors of black holes}\label{grey}
When quantum effects are taken into account, a black hole can emit thermal radiation as a blackbody spectrum at its event horizon called the Hawking radiation \cite{Hawking}. While propagating away from the black hole, the Hawking radiation encounters the black hole spacetime curvature, which modifies it. Therefore, an observer at a distant no longer sees the blackbody spectrum. Rather, he measures a modified one, which is called greybody. A quantity that measures how much the modified spectrum deviates from the blackbody spectrum is called the greybody factor. From a black hole scattering point of view, a black hole spacetime curvature acts as a potential barrier. The greybody factor is defined as a transmission coefficient.

There are various methods to compute the greybody factor. One of them is to find its lower bound rather than its exact value. The bounds of the greybody factors are given by \cite{Visser,Shabat}
\begin{equation}
T \geq \textrm{sech}^{2}\left[\int_{-\infty}^{\infty}\vartheta dr_{*}\right],
\end{equation}
where
\begin{equation}
\vartheta \equiv \frac{\sqrt{[h'(r_{*})]^{2} + [\omega^{2} - V(\tilde{r}) - h^{2}(r_{*})]^{2}}}{2h(r_{*})},
\end{equation}
for some positive function $h$ satisfying $h(-\infty)$ =

\noindent $\sqrt{\omega^{2} - V_{-\infty}}$ and $h(\infty) = \sqrt{\omega^{2} - V_{\infty}}$, where $V_{\pm\infty} = V(\pm\infty)$. For a potential satisfying $V_{\pm\infty} = 0$, we can set $h = \omega$, then
\begin{equation}
T \geq \textrm{sech}^{2}\left[\frac{1}{2\omega}\int_{-\infty}^{\infty} \vert V(\tilde{r})\vert dr_{*}\right].
\end{equation}
From equation (\ref{dr*}) and $\tilde{r} = r/c$, we obtain
\begin{equation}
T \geq \textrm{sech}^{2}\left[\frac{1}{2\tilde{\omega}}\int_{\tilde{r}_{1}}^{\tilde{r}_{2}}|\tilde{V}(\tilde{r})|\frac{1}{|f(\tilde{r})|} d\tilde{r}\right],\label{T}
\end{equation}
where $\tilde{\omega} = \omega c$. From the triangle inequality,
\begin{equation}
\left|\int_{a}^{b}f(x)dx\right| \leq \int_{a}^{b}|f(x)|dx,
\end{equation}
we obtain
\begin{equation}
\textrm{sech}\left(\int_{a}^{b}f(x)dx\right) \geq \textrm{sech}\left(\int_{a}^{b}|f(x)|dx\right),
\end{equation}
which follows from
\begin{equation}
\textrm{sech}(|x|) = \textrm{sech}(x).
\end{equation}
Then,
\begin{equation}
\textrm{sech}^{2}\left(\int_{a}^{b}f(x)dx\right) \geq \textrm{sech}^{2}\left(\int_{a}^{b}|f(x)|dx\right). \label{sech2}
\end{equation}
From equation (\ref{T}), we define
\begin{equation}
T_{\textrm{app}} \equiv \textrm{sech}^{2}\left[\frac{1}{2\tilde{\omega}}\int_{\tilde{r}_{1}}^{\tilde{r}_{2}}\frac{\tilde{V}(\tilde{r})}{f(\tilde{r})}d\tilde{r}\right] \label{Tapp}
\end{equation}
and
\begin{equation}
T_{\textrm{b}} \equiv \textrm{sech}^{2}\left[\frac{1}{2\tilde{\omega}}\int_{\tilde{r}_{1}}^{\tilde{r}_{2}}|\tilde{V}(\tilde{r})|\frac{1}{|f(\tilde{r})|} d\tilde{r}\right]. \label{Tb}
\end{equation}
From equation (\ref{sech2}), we obtain
\begin{equation}
T_{\textrm{app}} \geq T_{\textrm{b}}. \label{TappTb}
\end{equation}
We can rewrite equation (\ref{T}) as
\begin{equation}
T \geq T_{\textrm{b}}. \label{TTb}
\end{equation}
If $T \geq T_{\textrm{app}}$, we may say that $T_{\textrm{app}}$ is a better bound than $T_{\textrm{b}}$. If $T_{\textrm{app}} \geq T$, we cannot conclude which one, $T_{\textrm{app}}$ or $T_{\textrm{b}}$, is a better bound.

It is worthwhile to note that the calculation of the greybody factor by using the rigorous bound can provide us with a useful way to analyze the behaviour qualitatively, since it is possible to find the explicit expression for the greybody factor. However, since it is the bound, it may not provide the approximated value of the greybody factor. For example, in the high $\ell$ case, the bound is much lower than the approximated value. In this case, it is more convenient to use other methods to calculate the bound. In this paper,  we choose to investigate the greybody factor by using the third order of the WKB approximation proposed by Iyer and Will \cite{Iyer:1986np}. This is a well-developed method to study the barrier-like Quasi-normal modes and the greybody factor. 

For the WKB approximation, it is convenient to rewrite the redial equation in Eq.~(\ref{RE}) in the form as
\begin{equation}
\left(\frac{d^{2}}{dr_{*}^{2}}+Q\right)\psi=0,
\end{equation}
where $Q = \omega^{2} - V$. Moreover, we also use the  approximation as $\omega^{2} \approx V_{\max}$, where $V_{\max}$ denotes the maximum value of the effective potential. The greybody factor for this approximation is given by \cite{Cho:2004wj}
\begin{equation}\label{absor}
T=\frac{1}{1+\exp^{2S\left(\omega\right)}}.
\end{equation}
The function $S(\omega)$ can be expressed as
\begin{equation}
S(\omega) = \nonumber
\end{equation}
\begin{equation}\label{ss}
\begin{aligned}
& \pi k^{1/2}\left[\frac{1}{2}z_{0}^{2} + \left(\frac{15}{64}b^{2}_{3} - \frac{3}{16}b_{4} \right)z_{0}^{4} \right]\\
+ & \pi k^{-1/2}\left[\frac{3}{16}b_{4} -\frac{7}{64}b_{3}^{2} \right]\\
+ & \pi k ^{1/2}\left[\frac{1155}{2048}b_{3}^{4} - \frac{315}{256}b_{3}^{2}b_{4} + \frac{35}{128}b^{2}_{4}\right.\\
& \left. + \frac{35}{64}b_{3}b_{5} - \frac{5}{32}b_{6} \right]z_{0}^{6} \\
- & \pi k^{-1/2}\left[\frac{1365}{2048}b_{3}^{4} - \frac{525}{256}b_{3}^{2}b_{4} + \frac{85}{128}b_{4}^{2}\right.\\
& \left. + \frac{95}{64}b_{3}b_{5} - \frac{25}{32}b_{6} \right]z_{0}^{2} + O\left(\omega\right),
\end{aligned}
\end{equation}
where $O\left(\omega\right)$ is a set of the higher order terms and
\begin{equation}\label{sscoes}
\begin{aligned}
z_{0}^{2} & = -\frac{Q_{\max}}{k}\\
k         & = \frac{1}{2}\left(\frac{d^{2}Q}{dr_{*}^{2}} \right)_{\max}\\
b_{n}     & = \left(\frac{1}{n!k} \right)\left(\frac{d^{n}Q}{dr_{*}^{n}} \right)_{\max}.
\end{aligned}
\end{equation}
Note that the subscript ``$\max$" denotes the quantities for $r = r_{\max}$ after taking the derivative where $r_{\max}$ denotes the position of $V_{\max}$. In this way, one can find the greybody factor by using the numerical method. Note that the results evaluated by using the WKB method cannot be done for low multipole, especially for $\ell =0$. In this case, we cannot compare the results between such two methods.

\subsection{Greybody factors for the dS case}
A dS black hole has two horizons; namely, event horizon and the cosmological horizon. Its potential is shown in Figure \ref{vdsdrgt}. It can be seen that the dS potential for $\ell = 0$ with $\tilde{m} < \tilde{m}_{\min}$ can be both positive and negative, while one for $\ell = 0$ with $\tilde{m} > \tilde{m}_{\min}$ and $\ell = 1$ is always positive. Therefore, finding the bounds on the greybody factors for such two cases can be done independently.

\subsubsection{$\ell = 0$ and $\tilde{m} < \tilde{m}_{\min}$}
\begin{figure}[H]
\begin{subfigure}{.45\textwidth}
  \centering
  \includegraphics[width=0.8\linewidth]{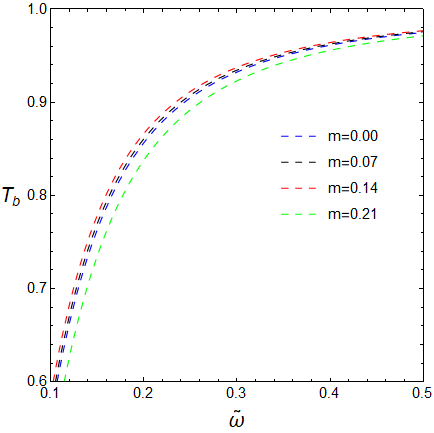}
  \caption{}
\end{subfigure}
\begin{subfigure}{.45\textwidth}
\centering
\includegraphics[width=0.8\linewidth]{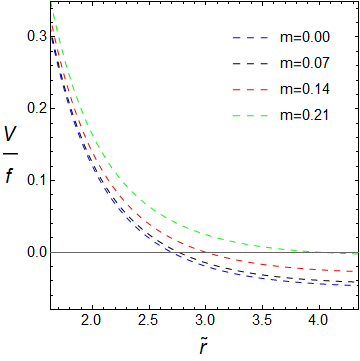}
  \caption{}
\end{subfigure}
\caption{\label{bound}(a) The rigorous bounds on the greybody factors in the dS case with $\ell = 0$, $\tilde{M} = 1$, $\alpha_{g} = 1$, $c_{2} = -0.01$, $\beta_{m} = 0.875$ and $\tilde{m} = 0$, 0.07, 0.14 and 0.21. (b) The graph of the function $\tilde{V}/f$ in the dS case with the same parameters.}
\end{figure}

For $\ell = 0$, the dS potential is positive for $\tilde{r}_{1} < \tilde{r} < \tilde{r}_{\textrm{cri}}$ and negative for $\tilde{r}_{\textrm{cri}} < \tilde{r} < \tilde{r}_{2}$, while $f_{\textrm{dS}}(\tilde{r})$ is always positive for $\tilde{r}_{1} < \tilde{r} < \tilde{r}_{2}$, where $\tilde{r}_{\textrm{cri}}$ is the point satisfying $f'_{\textrm{dS}}(\tilde{r}_{\textrm{cri}}) = -\tilde{m}^{2}\tilde{r}_{\textrm{cri}}$. We can write
\begin{equation}
T_{\textrm{dS}, \ell = 0} \geq T_{\textrm{b, dS}, \ell = 0},
\end{equation}
where
\begin{eqnarray}
T_{\textrm{b, dS}, \ell = 0} &=& \textrm{sech}^{2}\left[\frac{1}{2\tilde{\omega}}\int_{\tilde{r}_{1}}^{\tilde{r}_{\textrm{cri}}}\frac{\tilde{V}_{\textrm{dS}, \ell = 0}(\tilde{r})}{f_{\textrm{dS}}(\tilde{r})}d\tilde{r}\right.\nonumber\\
 && \left. - \frac{1}{2\tilde{\omega}}\int_{\tilde{r}_{\textrm{cri}}}^{\tilde{r}_{2}}\frac{\tilde{V}_{\textrm{dS}, \ell = 0}(\tilde{r})}{f_{\textrm{dS}}(\tilde{r})}d\tilde{r}\right]
\end{eqnarray}
and, from equation (\ref{Vds}), $\tilde{V}_{\textrm{dS}, \ell = 0}(\tilde{r})/f_{\textrm{dS}}(\tilde{r})$ is given by
\begin{equation}
\frac{\tilde{V}_{\textrm{dS}, \ell = 0}(\tilde{r})}{f_{\textrm{dS}}(\tilde{r})} = \frac{2\tilde{M}}{\tilde{r}^{3}} + 2\alpha_{g}c_{2} - \frac{3\alpha_{g}\sqrt[3]{4c_{2}^{2}}}{\tilde{r}} + \tilde{m}^{2}.
\end{equation}
Calculating the integrals, we obtain
\begin{eqnarray}
T_{\textrm{b, dS}, \ell = 0} &=& \textrm{sech}^{2}\left[\frac{1}{2\tilde{\omega}}\left\{\tilde{M}\left(\frac{1}{\tilde{r}_{1}^{2}} + \frac{1}{\tilde{r}_{2}^{2}} - \frac{2}{\tilde{r}_{\textrm{cri}}^{2}}\right)\right.\right.\nonumber\\
      && - 2\alpha_{g}c_{2}(\tilde{r}_{1} + \tilde{r}_{2} - 2\tilde{r}_{\textrm{cri}})\nonumber\\
      && - 3\alpha_{g}\sqrt[3]{4c_{2}^{2}}\ln\frac{\tilde{r}_{\textrm{cri}}^{2}}{\tilde{r}_{1}\tilde{r}_{2}}\nonumber\\
      && \left.\left. - \tilde{m}^{2}(\tilde{r}_{1} + \tilde{r}_{2} - 2\tilde{r}_{\textrm{cri}})\right\}\right]. \label{Tbdsl0}
\end{eqnarray}
Since the bound of the greybody factor depends on the integral of $\tilde{V}/f$, it can be analyzed by determining the absolute area under the graph of $\tilde{V}/f$. In this case, there exists the negative part of $\tilde{V}/f$, so that it is possible to minimize the area by choosing the proper value of $\tilde{m}$. This corresponds to the  maximized greybody factor bound $T_{\textrm{b}}$. In fact, one can find $\tilde{r}$ in terms of $\tilde{m}$ from $\tilde{V}/f|_{\tilde{r} = \tilde{r}_{\textrm{cri}}} = 0$. We can find the integral of $\tilde{V}/f$, which depends on $\tilde{r}_{\textrm{cri}}$. Therefore, in principle, one can find the integral, which is a function of $\tilde{m}$. Then, we can find the critical mass $\tilde{m}_{\textrm{c}}$ to minimize the integral corresponding to the maximized $T_{\textrm{b}}$. The expression is very lengthy, we do not express it here. For $c_{2} = -0.01, \beta_{m} = 0.875$, the critical mass can be evaluated as $\tilde{m}_{\textrm{c}} \sim 0.14$. This behaviour of the bounds on the greybody factors is shown in Figure \ref{bound}. From Figure \ref{bound}(a), we see that the bound on the greybody factors increases with increasing $\tilde{m}$ for $0 \leq \tilde{m} \leq 0.14$. For $\tilde{m} > 0.14$ the bound starts to decrease with increasing $\tilde{m}$. This can be inferred from Figure \ref{bound}(b) that the absolute of the area under the graph on an interval $\tilde{r}_{1} < \tilde{r} < \tilde{r}_{\textrm{cri}}$ increases with increasing $\tilde{m}$ and that on an interval $\tilde{r}_{\textrm{cri}} < \tilde{r} < \tilde{r}_{2}$ decreases with increasing $\tilde{m}$. As a result, there exists a critical mass of the scalar field, which provides the maximum bound of the greybody factor. This is a crucial behaviour of the scalar field mass on the greybody factor.

\begin{figure}[H]
\begin{subfigure}{.45\textwidth}
  \centering
  \includegraphics[width=0.8\linewidth]{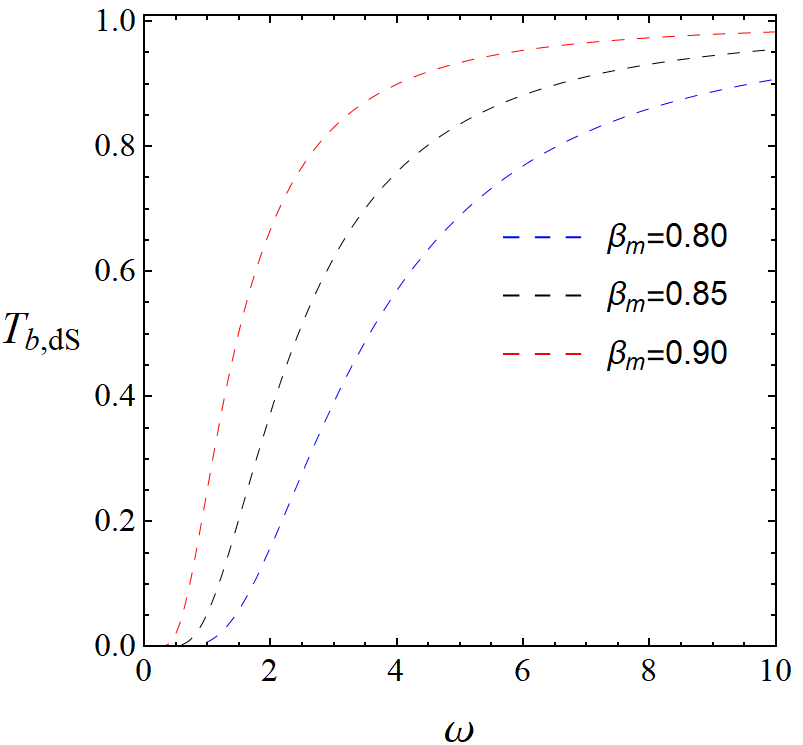}
  \caption{}
\end{subfigure}
\begin{subfigure}{.45\textwidth}
  \centering
  \includegraphics[width=0.8\linewidth]{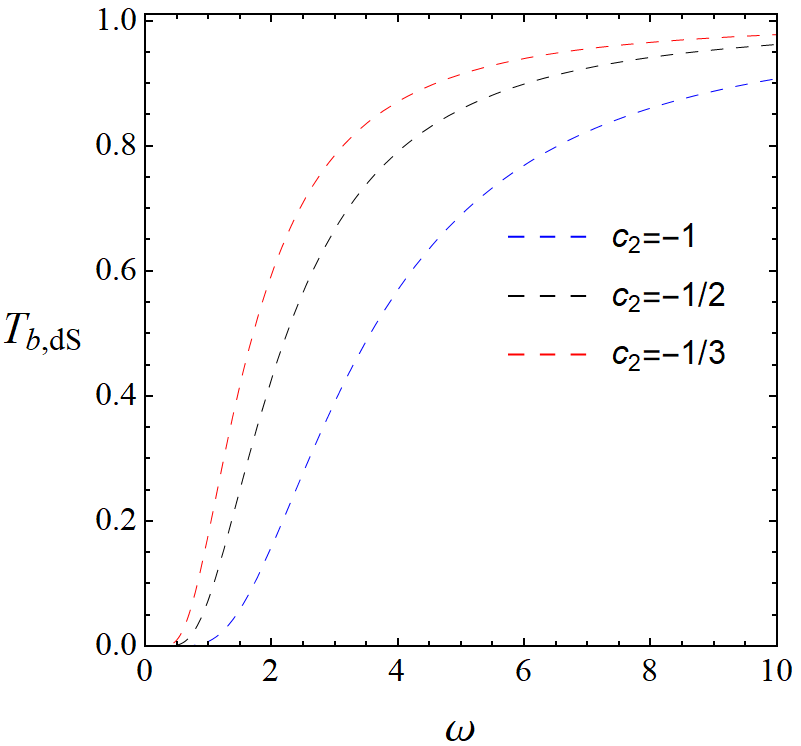}
  \caption{}
\end{subfigure}
\caption{\label{vf}The rigorous bounds on the greybody factors in the dS case with $\ell = 0$, $\tilde{M} = 1$, $\alpha_{g} = 1$. (a) $c_{2} = -1$, $\tilde{m} = 0.2$ and $\beta_{m} = 0.8$, 0.85 and 0.9. (b) $\tilde{m} = 0.2$, $\beta_{m} = 0.8$ and $c_{2} = -1$, -1/2 and -1/3.}
\end{figure}

The bounds on the greybody factors with different $\beta_{m}$ and $c_{2}$ are plotted in Figure \ref{vf}(a) and \ref{vf}(b), respectively. For Figure \ref{vf}(a), we see that when $\beta_{m}$ increases, the dS bound on the greybody factors also increases. This follows from Figure \ref{pot2}(a) where the local maximum of the potential decreases with increasing $\beta_{m}$.

For Figure \ref{vf}(b), we see that when $c_{2}$ decreases, corresponding to the increase in the magnitude of the positive cosmological constant, the bound on the greybody factors also decreases. This follows from Figure \ref{pot2}(b) where the local maximum of the potential increases with increasing $c_{2}$. In the background with the stronger cosmological constant, the massive scalar field can penetrate the dS potential, but with more difficulty.

\begin{figure}[H]
\begin{subfigure}{.5\textwidth}
  \centering
  \includegraphics[width=0.7\linewidth]{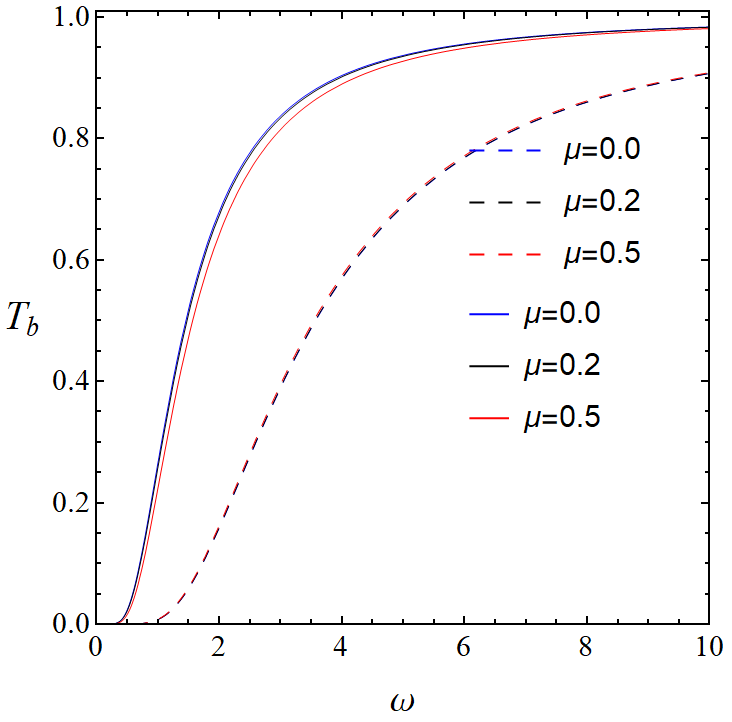}
  \caption{}
\end{subfigure}
\begin{subfigure}{.5\textwidth}
  \centering
  \includegraphics[width=0.7\linewidth]{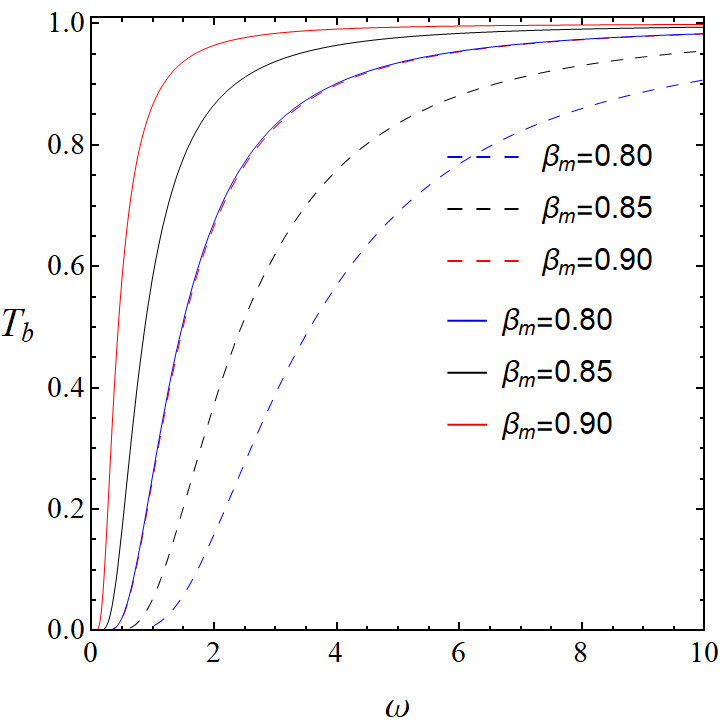}
  \caption{}
\end{subfigure}
\begin{subfigure}{.5\textwidth}
  \centering
  \includegraphics[width=0.7\linewidth]{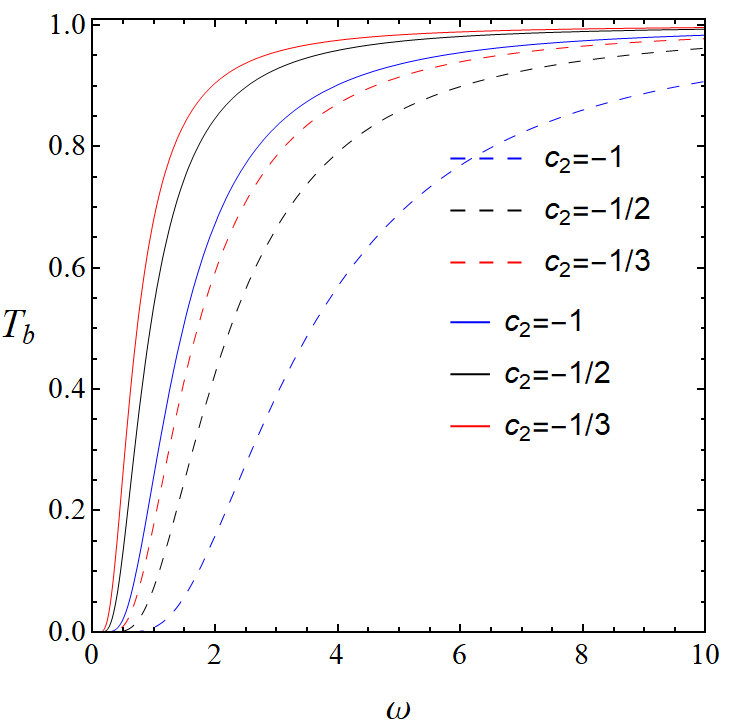}
  \caption{}
\end{subfigure}
\caption{\label{bound1}Comparison of $T_{\textrm{app}}$ and $T_{\textrm{b}}$ defined in equations (\ref{Tapp}) and (\ref{Tb}), respectively, in the dS case.}
\end{figure}

The graphs of $T_{\textrm{app}}$ with different $\tilde{m}$, $\beta_{m}$ and $c_{2}$ are plotted in Figure \ref{bound1}(a), \ref{bound1}(b) and \ref{bound1}(c), respectively. From Figure \ref{bound1}(a), we see that when $\tilde{m}$ increases, the bound on the greybody factors decreases because the result follows from the dS potential in Figure \ref{pot}, which increases with increasing $\tilde{m}$. The more massive the scalar field, it can be harder to transmit out from the dS potential.

For Figure \ref{bound1}(b), we see that when $\beta_{m}$ increases, the dS bound on the greybody factors also increases. This follows from Figure \ref{pot2}(a) where the local maximum of the potential decreases with increasing $\beta_{m}$.

For Figure \ref{bound1}(c), we see that when $c_{2}$ decreases, corresponding to increase in the magnitude of the positive cosmological constant, the bound on the greybody factors also decreases. This follows from Figure \ref{pot2}(b) where the local maximum of the potential increases with increasing $c_{2}$. In the background with the stronger cosmological constant, the massive scalar field can penetrate the dS potential, but with more difficulty.

Moreover, we explicitly see that $T_{\textrm{app}}$ (the solid curve) is greater than or equal to $T_{\textrm{b}}$ (the dash curve) according to the inequality (\ref{TappTb}).

\subsubsection{$\ell = 0$ with $\tilde{m} > \tilde{m}_{\min}$ and $\ell = 1$}
\begin{figure}[H]
\begin{subfigure}{.5\textwidth}
  \centering
\includegraphics[width=0.8\linewidth]{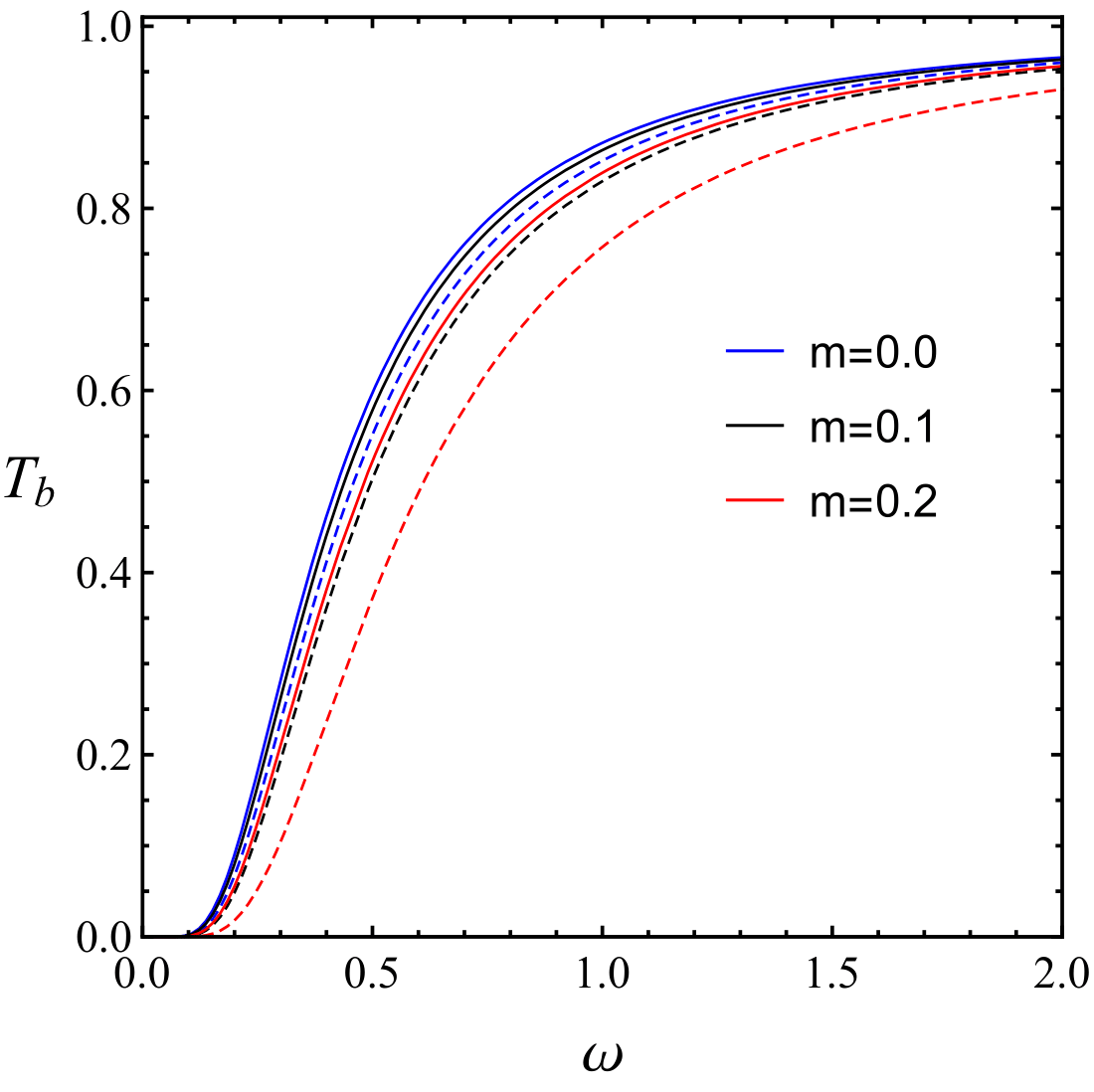}
  \caption{}
\end{subfigure}
\begin{subfigure}{.5\textwidth}
  \centering
 \includegraphics[width=0.8\linewidth]{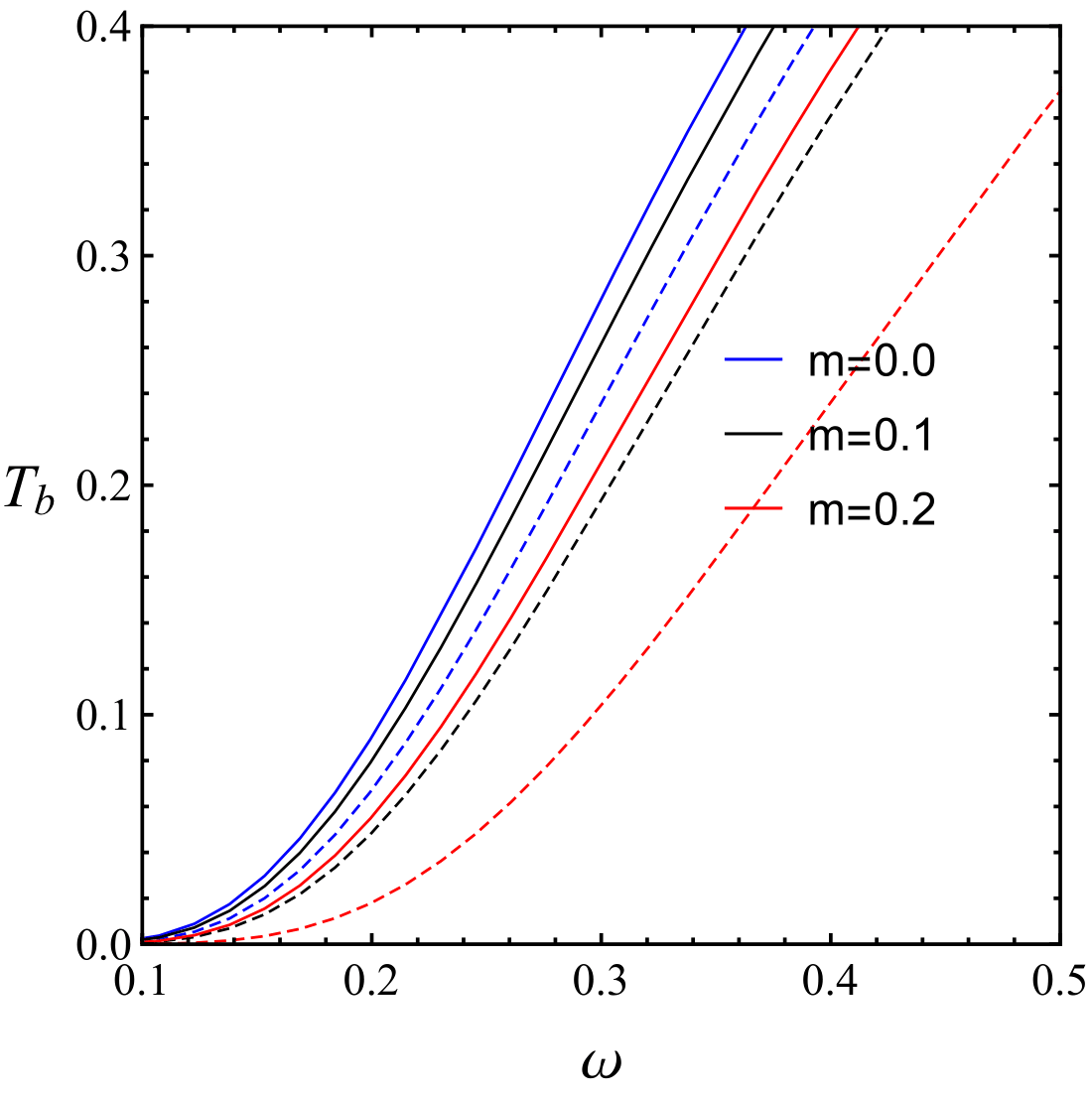}
  \caption{}
\end{subfigure}
\caption{\label{Tbmg-TdS}(a) The bounds on the greybody factors of the SdS black holes (dashed line) and the asymptotically dS black holes in the dRGT massive gravity (solid line) for $c_{2} = -0.01$, $\beta_{m} = 0.89$, $\ell = 1$ and $\tilde{m} = 0$, 0.1 and 0.2. (b) The magnification of (a).}
\end{figure}

From Figure \ref{vdsdrgt}, the potential for $\ell = 0$ with $\tilde{m} > \tilde{m}_{\min}$ and $\ell = 1$ is positive on the interval $\tilde{r}_{1} < \tilde{r} < \tilde{r}_{2}$. Therefore, from equations (\ref{Tapp}) and (\ref{Tb}), $T_{\textrm{b}} = T_{\textrm{app}}$. Then, equation (\ref{TTb}) becomes
\begin{equation}
T_{\textrm{dS}} \geq T_{\textrm{b}, \textrm{dS}},
\end{equation}
where
\begin{equation}
T_{\textrm{b}, \textrm{dS}} = \textrm{sech}^{2}\left[\frac{1}{2\tilde{\omega}}\int_{\tilde{r}_{1}}^{\tilde{r}_{2}}\frac{\tilde{V}_{\textrm{dS}}(\tilde{r})}{f_{\textrm{dS}}(\tilde{r})}d\tilde{r}\right]
\end{equation}
and
\begin{equation}
\frac{\tilde{V}_{\textrm{dS}}(\tilde{r})}{f_{\textrm{dS}}(\tilde{r})} = \frac{\ell(\ell + 1)}{\tilde{r}^{2}} + \frac{2\tilde{M}}{\tilde{r}^{3}} + 2\alpha_{g}c_{2} - \frac{3\alpha_{g}\sqrt[3]{4c_{2}^{2}}}{\tilde{r}} + \tilde{m}^{2}.
\end{equation}
Calculating the integrals, we obtain
\begin{eqnarray}
T_{\textrm{b}, \textrm{dS}} &=& \textrm{sech}^{2}\left[\frac{1}{2\tilde{\omega}}\left\{\ell(\ell + 1)\left(\frac{1}{\tilde{r}_{1}} - \frac{1}{\tilde{r}_{2}}\right)\right.\right.\nonumber\\
      && + \tilde{M}\left(\frac{1}{\tilde{r}_{1}^{2}} - \frac{1}{\tilde{r}_{2}^{2}}\right) + 2\alpha_{g}c_{2}(\tilde{r}_{2} - \tilde{r}_{1})\nonumber\\
      && \left.\left. - 3\alpha_{g}\sqrt[3]{4c_{2}^{2}}\ln\frac{\tilde{r}_{2}}{\tilde{r}_{1}} + \tilde{m}^{2}(\tilde{r}_{2} - \tilde{r}_{1})\right\}\right]. \label{Tbdsl1}
\end{eqnarray}
For SdS black holes, the logarithmic term disappears and the above bound reduces to
\begin{eqnarray}
T_{\textrm{b}, \textrm{SdS}} &=& \textrm{sech}^{2}\left[\frac{1}{2\tilde{\omega}}\left\{\ell(\ell + 1)\left(\frac{1}{\tilde{r}_{1}} - \frac{1}{\tilde{r}_{2}}\right)\right.\right.\nonumber\\
      && + \tilde{M}\left(\frac{1}{\tilde{r}_{1}^{2}} - \frac{1}{\tilde{r}_{2}^{2}}\right) + 2\alpha_{g}c_{2}(\tilde{r}_{2} - \tilde{r}_{1})\nonumber\\
      && \left.\left. + \tilde{m}^{2}(\tilde{r}_{2} - \tilde{r}_{1})\right\}\right].
\end{eqnarray}
The bounds on the greybody factors of the SdS black holes and the asymptotically dS black holes in the dRGT massive gravity are plotted in Figure \ref{Tbmg-TdS}. We see that the bounds of the SdS black holes are lower than ones of the asymptotically dS black holes in the dRGT massive gravity. This follows from the fact that the logarithmic term in equation (\ref{Tbdsl1}) is negative. Therefore, $T_{\textrm{b}, \textrm{SdS}} < T_{\textrm{b}, \textrm{dS}}$ because sech($x$) is a decreasing function for $x > 0$.

Moreover, when $m$ increases, both $T_{\textrm{b}, \textrm{SdS}}$ and $T_{\textrm{b}, \textrm{dS}}$ decreases. This is apparent because again sech($x$) is a decreasing function for $x > 0$.

\begin{figure}[H]
\begin{subfigure}{.45\textwidth}
  \centering
\includegraphics[width=0.8\linewidth]{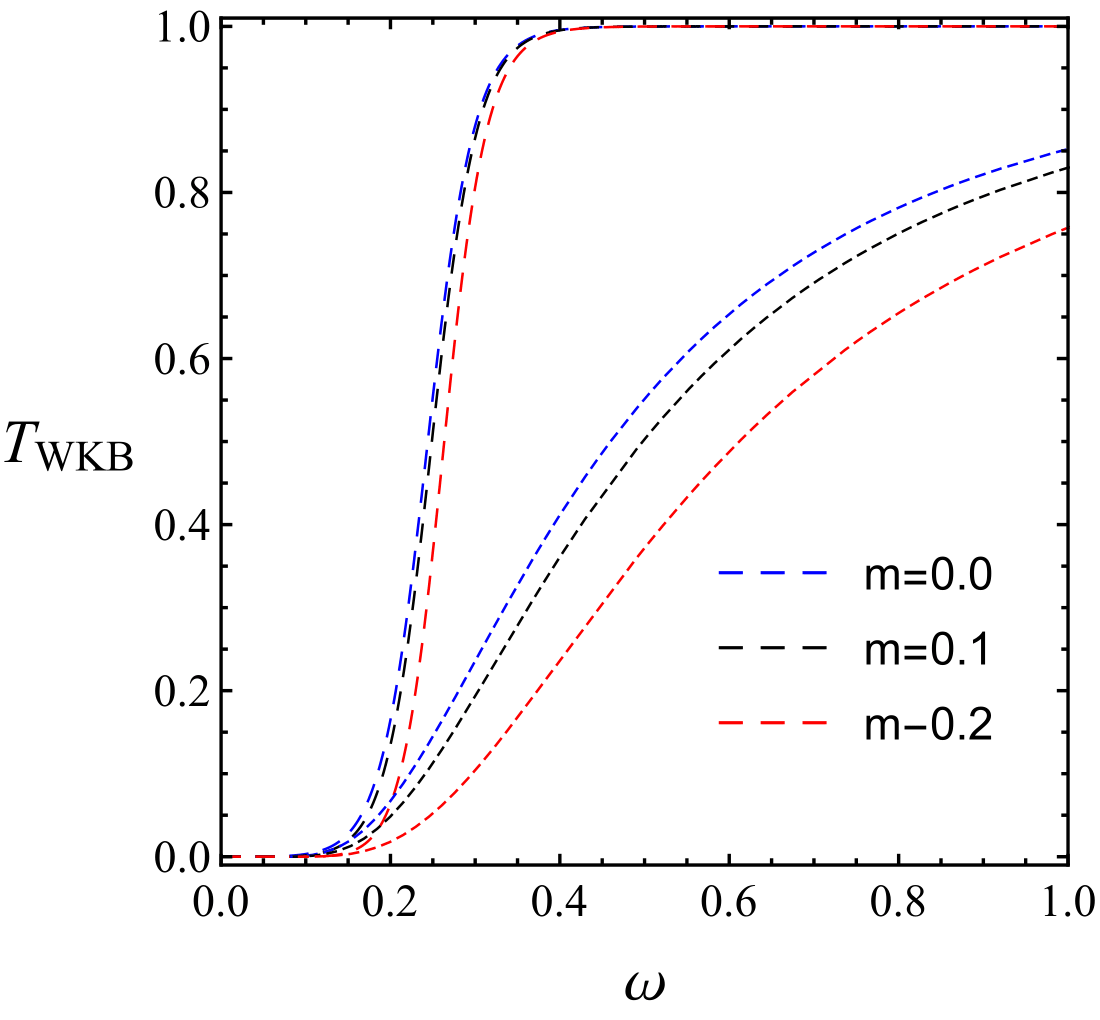}
  \caption{}
\end{subfigure}
\begin{subfigure}{.45\textwidth}
  \centering
 \includegraphics[width=0.8\linewidth]{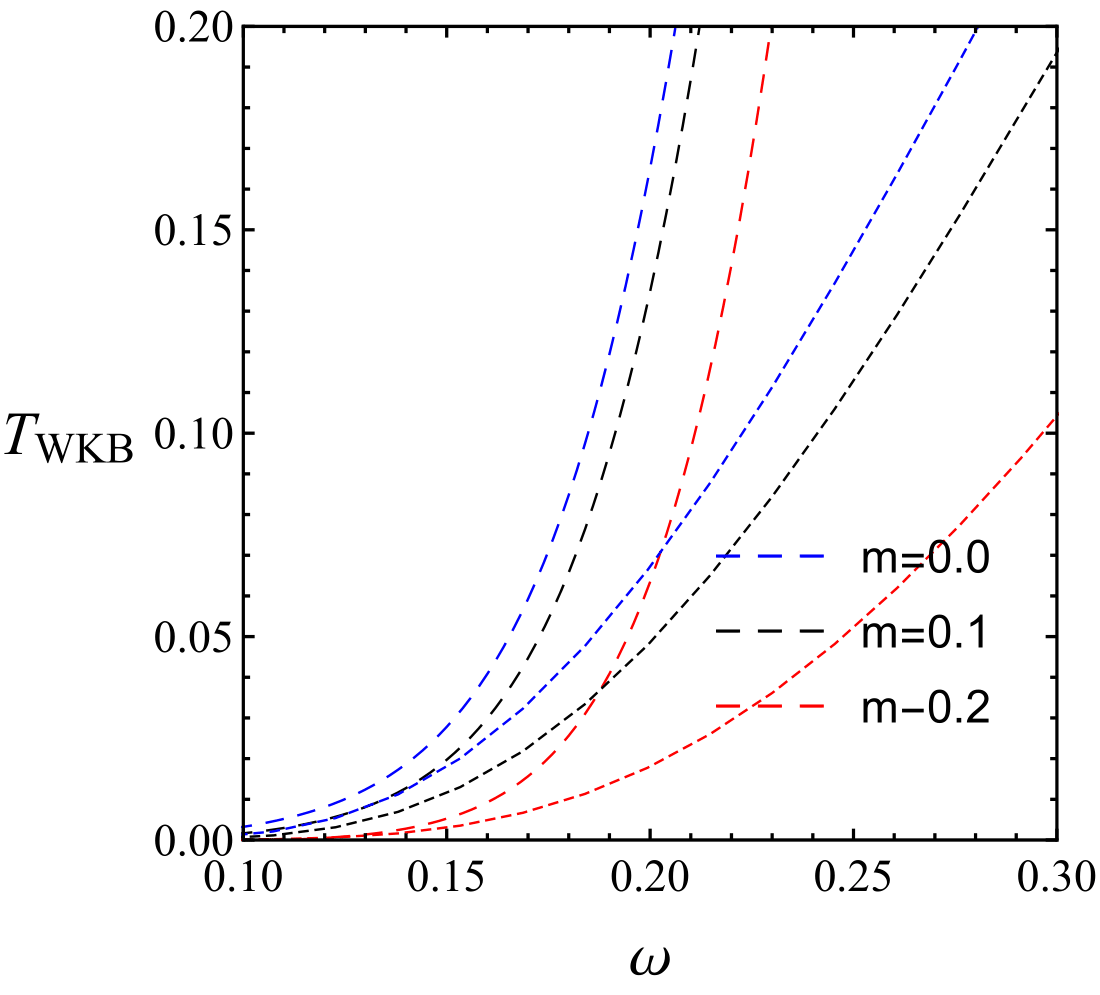}
  \caption{}
\end{subfigure}
\caption{\label{TWKB-TdS}(a) The greybody factors of the SdS black holes $T_{\textrm{b}, \textrm{SdS}, \ell = 1}$ from the bound (dotted line) and $T_{\textrm{WKB}}$ from the WKB method (dashed line) for $c_{2} = -0.01$, $\ell = 1$ and $\tilde{m} = 0$, 0.1 and 0.2. (b) The magnification of (a).}
\end{figure}

The results from the WKB approximation can be used  for $\ell \neq 0$ as we mentioned above.
The graphs of $T_{\textrm{WKB}}$ compared to $T_{\textrm{b}, \textrm{SdS}, \ell = 1}$ with different $m$ for the SdS black holes are plotted in Figure \ref{TWKB-TdS}. We see that $T_{\textrm{b}, \textrm{SdS}, \ell = 1} \leq T_{\textrm{WKB}}$, which indicates that $T_{\textrm{b}, \textrm{SdS}, \ell = 1}$ is a true lower bound. Moreover, $T_{\textrm{WKB}}$ and $T_{\textrm{b}, \textrm{SdS}, \ell = 1}$ are so close to one another when $\tilde{\omega}$ is small.

\begin{figure}[H]
\begin{subfigure}{.45\textwidth}
  \centering
\includegraphics[width=0.8\linewidth]{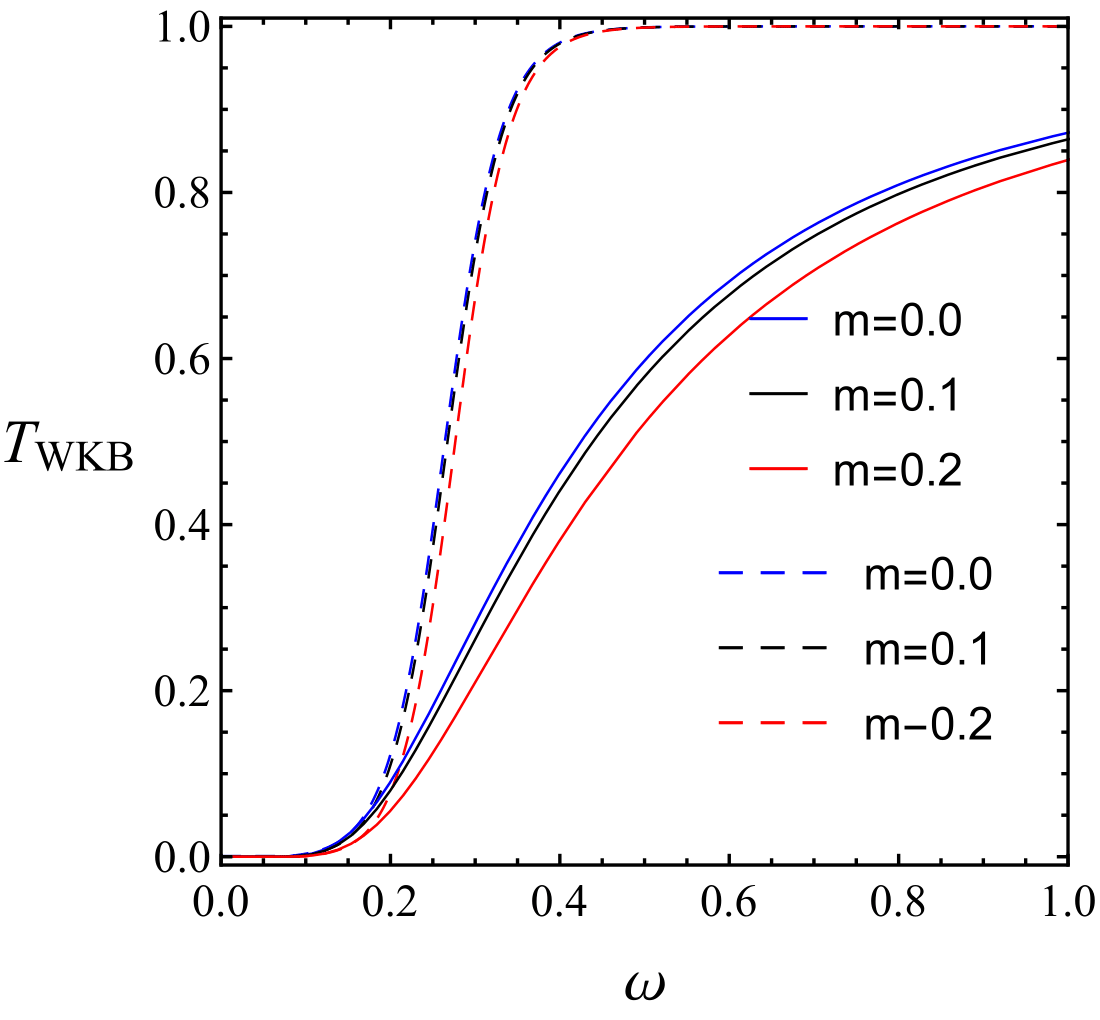}
  \caption{}
\end{subfigure}
\begin{subfigure}{.45\textwidth}
  \centering
 \includegraphics[width=0.8\linewidth]{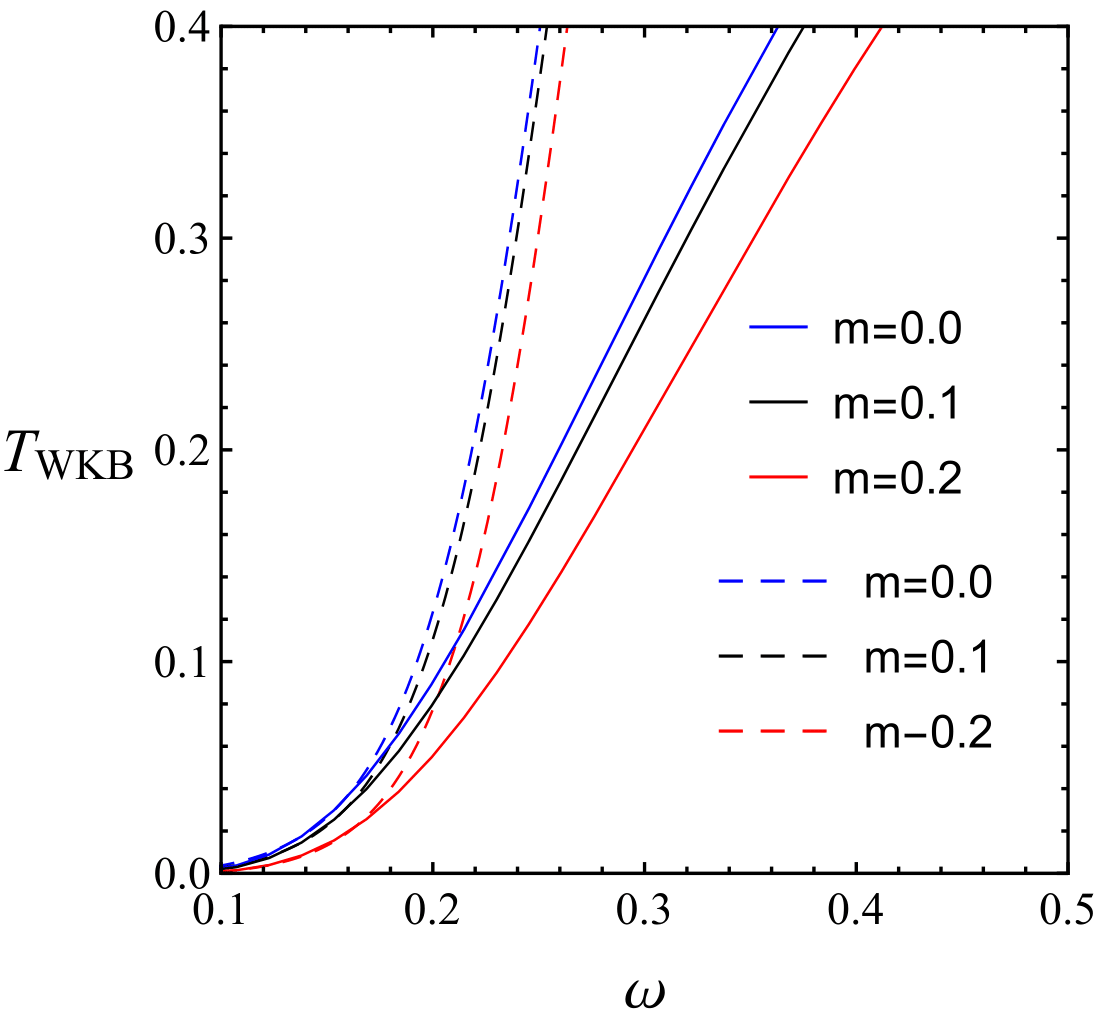}
  \caption{}
\end{subfigure}
\caption{\label{TWKB-TdS2}(a) The greybody factors of the asymptotically dS black holes in the dRGT massive gravity $T_{\textrm{b}, \textrm{dS}, \ell = 1}$ from the bound (solid line) and $T_{\textrm{WKB}}$ from the WKB method (dashed line) for $c_{2} = -0.01$, $\beta_{m} = 0.89$, $\ell = 1$ and $\tilde{m} = 0$, 0.1 and 0.2. (b) The magnification of (a).}
\end{figure}

The graphs of $T_{\textrm{WKB}}$ compared to $T_{\textrm{b}, \textrm{dS}, \ell = 1}$ with different $\tilde{m}$ for the asymptotically dS black holes in the dRGT massive gravity are plotted in Figure \ref{TWKB-TdS2}. We see that $T_{\textrm{b}, \textrm{dS}, \ell = 1} \leq T_{\textrm{WKB}}$, which indicates that $T_{\textrm{b}, \textrm{dS}, \ell = 1}$ is also a true lower bound. Moreover, $T_{\textrm{WKB}}$ and $T_{\textrm{b}, \textrm{dS}, \ell = 1}$ are so close to one another when $\tilde{\omega}$ is small.

\subsection{Greybody factors of the AdS black hole}
An AdS black hole in the dRGT massive gravity can have more than one horizon, while a Schwarzschild AdS black hole has only one horizon. In this work, we are interested in the case where the AdS black hole has three horizons. Its potential is shown in Figure \ref{potadsl12}, where the AdS potentials for $\ell = 0$ can be both positive and negative on $\tilde{r}_{1} < \tilde{r} < \tilde{r}_{2}$ and $\tilde{r}_{2} < \tilde{r} < \tilde{r}_{3}$, while one for $\ell \neq 0$ is positive on $\tilde{r}_{1} < \tilde{r} < \tilde{r}_{2}$ and negative on $\tilde{r}_{2} < \tilde{r} < \tilde{r}_{3}$. Therefore, finding the bounds on the greybody factors for such two cases can be done independently.

\subsubsection{$\ell = 0$}
\begin{figure}[H]
\begin{subfigure}{.5\textwidth}
  \centering
  \includegraphics[width=0.5\linewidth]{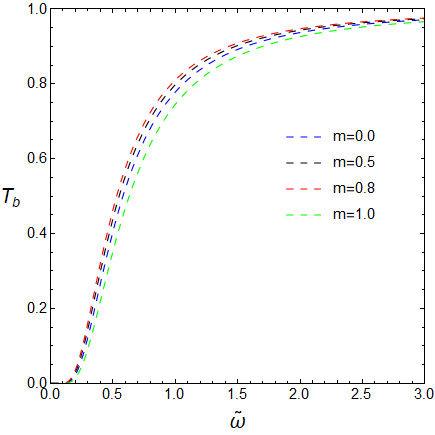}
  \caption{}
\end{subfigure}
\begin{subfigure}{.5\textwidth}
  \centering
  \includegraphics[width=0.5\linewidth]{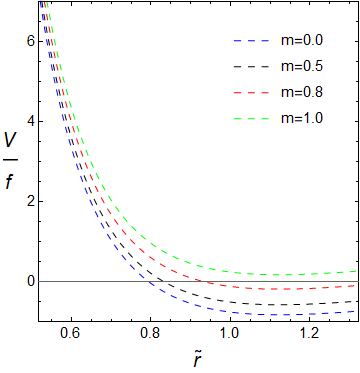}
  \caption{}
\end{subfigure}
\begin{subfigure}{.5\textwidth}
  \centering
  \includegraphics[width=0.5\linewidth]{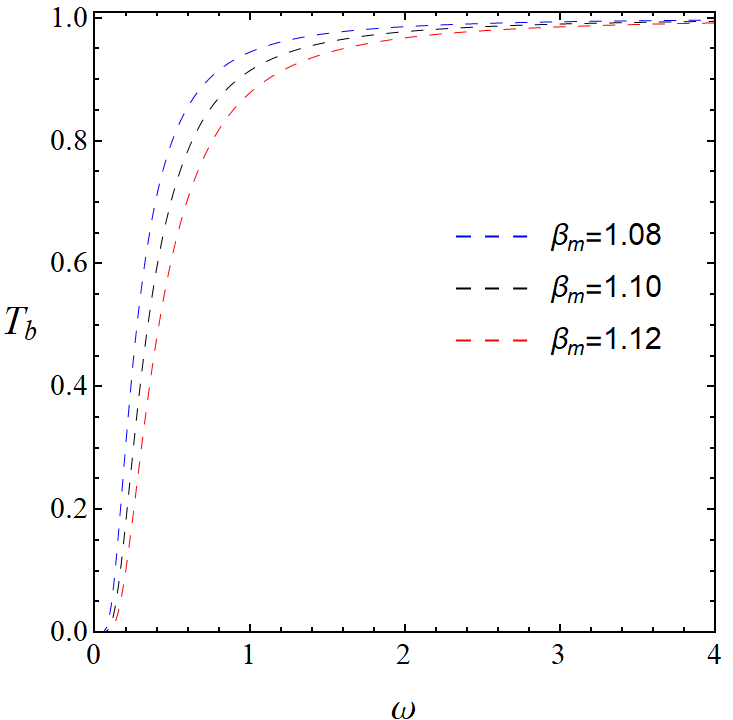}
  \caption{}
\end{subfigure}
\begin{subfigure}{.5\textwidth}
  \centering
  \includegraphics[width=0.5\linewidth]{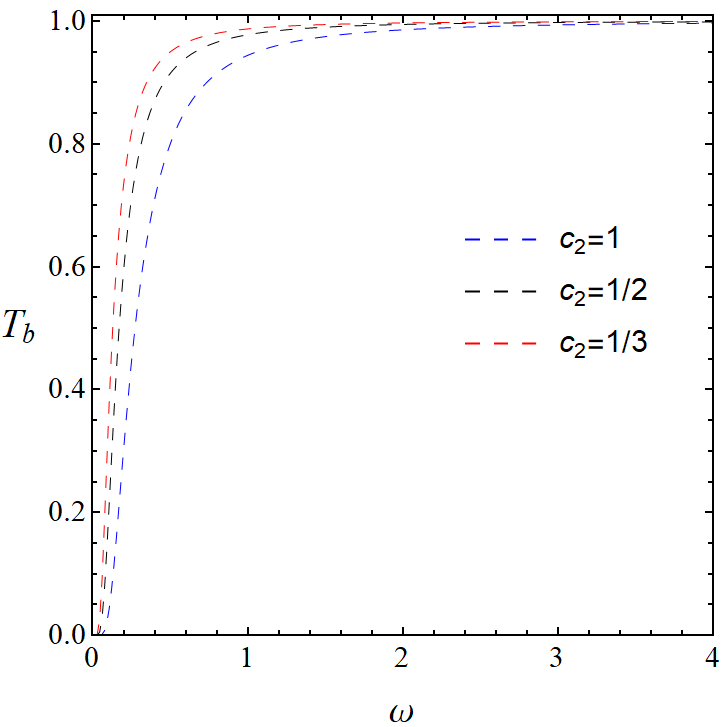}
  \caption{}
\end{subfigure}
\caption{\label{bound3}The rigorous bounds (between $\tilde{r}_{1}$ and $\tilde{r}_{2}$) on the greybody factors of the AdS black holes with $\ell = 0$, $\tilde{M} = 1$, $\alpha_{g} = 1$. (a) $c_{2} = 1$, $\beta_{m} = 1.08$ and $\tilde{m} = 0$, 0.2 and 0.5. (b) The magnification of (a). (c) $c_{2} = 1$, $\tilde{m} = 0.2$ and $\beta_{m} = 1.08$, 1.10 and 1.12. (d) $\tilde{m} = 0.2$, $\beta_{m} = 1.08$ and $c_{2} = 1$, 1/2 and 1/3.}
\end{figure}

We divide the region of $\tilde{r}$ into two areas, $\tilde{r}_{1} < \tilde{r} < \tilde{r}_{2}$ and $\tilde{r}_{2} < \tilde{r} < \tilde{r}_{3}$. In the first region, $f_{\textrm{AdS}}(\tilde{r})$ is always positive. However, the AdS potential is positive for $\tilde{r}_{1} < \tilde{r} < \tilde{r}_{\textrm{1cri}}$ and negative for $\tilde{r}_{\textrm{1cri}} < \tilde{r} < \tilde{r}_{2}$, where $\tilde{r}_{\textrm{1cri}}$ is the point satisfying $f'_{\textrm{AdS}}(\tilde{r}_{\textrm{1cri}}) = -\tilde{m}^{2}\tilde{r}_{\textrm{1cri}}$. From equation (\ref{T}), we can write
\begin{equation}
T_{\textrm{AdS}, \ell = 0} \geq T_{\textrm{b, AdS}, \ell = 0},
\end{equation}
where
\begin{eqnarray}
T_{\textrm{b, AdS}, \ell = 0} &=& \textrm{sech}^{2}\left[\frac{1}{2\tilde{\omega}}\int_{\tilde{r}_{1}}^{\tilde{r}_{\textrm{1cri}}}\frac{\tilde{V}_{\textrm{AdS}, \ell = 0}(\tilde{r})}{f_{\textrm{AdS}}(\tilde{r})}d\tilde{r}\right.\nonumber\\
    && \left. - \frac{1}{2\tilde{\omega}}\int_{\tilde{r}_{\textrm{1cri}}}^{\tilde{r}_{2}}\frac{\tilde{V}_{\textrm{AdS}, \ell = 0}(\tilde{r})}{f_{\textrm{AdS}}(\tilde{r})}d\tilde{r}\right]
\end{eqnarray}
and $\tilde{V}_{\textrm{AdS}, \ell = 0}(\tilde{r})/f_{\textrm{AdS}}(\tilde{r})$ is
\begin{equation}
\frac{\tilde{V}_{\textrm{AdS}, \ell = 0}(\tilde{r})}{f_{\textrm{AdS}}(\tilde{r})} = \frac{2\tilde{M}}{\tilde{r}^{3}} + 2\alpha_{g}c_{2} - \frac{3\alpha_{g}\sqrt[3]{4c_{2}^{2}}}{\tilde{r}} + \tilde{m}^{2}.
\end{equation}
Calculating the integrals, we obtain
\begin{eqnarray}
T_{\textrm{b, AdS}, \ell = 0} &=& \textrm{sech}^{2}\left[\frac{1}{2\tilde{\omega}}\left\{\tilde{M}\left(\frac{1}{\tilde{r}_{1}^{2}} + \frac{1}{\tilde{r}_{2}^{2}} - \frac{2}{\tilde{r}_{\textrm{1cri}}^{2}}\right)\right.\right.\nonumber\\
      && - 2\alpha_{g}c_{2}(\tilde{r}_{1} + \tilde{r}_{2} - 2\tilde{r}_{\textrm{1cri}})\nonumber\\
      && - 3\alpha_{g}\sqrt[3]{4c_{2}^{2}}\ln\frac{\tilde{r}_{\textrm{1cri}}^{2}}{\tilde{r}_{1}\tilde{r}_{2}}\nonumber\\
      && \left.\left. - \tilde{m}^{2}(\tilde{r}_{1} + \tilde{r}_{2} - 2\tilde{r}_{\textrm{1cri}})\right\}\right].\label{Tmads}
\end{eqnarray}
The bounds on the greybody factors with different $\tilde{m}$, $\beta_{m}$ and $c_{2}$ are plotted in Figure \ref{bound3}(a), \ref{bound3}(c) and \ref{bound3}(d), respectively. In Figure \ref{bound3}(a), we see  when the scalar mass increases, the greybody factor bound keeps increasing until $\tilde{m}=0.8$ and then decreasing for $\tilde{m}>0.8$. In fact, by performing in the same way as done in the dS case, we found that there exists a critical mass in which the greybody bound is maximum. For example, by setting $\tilde{M} = 1$, $\alpha_{g} = 1$, $c_{2} = 1$, and  $\beta_{m} = 1.08$, we found that the critical mass is at $m_c = 0.7937$. This can be seen by considering the area under the function $|V/f|$ as found in Figure \ref{bound3}(b).

For Figure \ref{bound3}(c), we see that when $\beta_{m}$ increases, the bound on the greybody factors decreases. This is interesting because from Figure \ref{pot4}(a), the local maximum of the potential also decreases with increasing $\beta_{m}$. The changes in the AdS potential and the AdS bound with $\beta_{m}$ are the same.

For Figure \ref{bound3}(d), we see that when $c_{2}$ increases, the bound on the greybody factors decreases. For the background of the stronger cosmological constant, the massive scalar field can be harder to transmit from the AdS potential.

\begin{figure}[H]
\begin{subfigure}{.5\textwidth}
  \centering
  \includegraphics[width=0.7\linewidth]{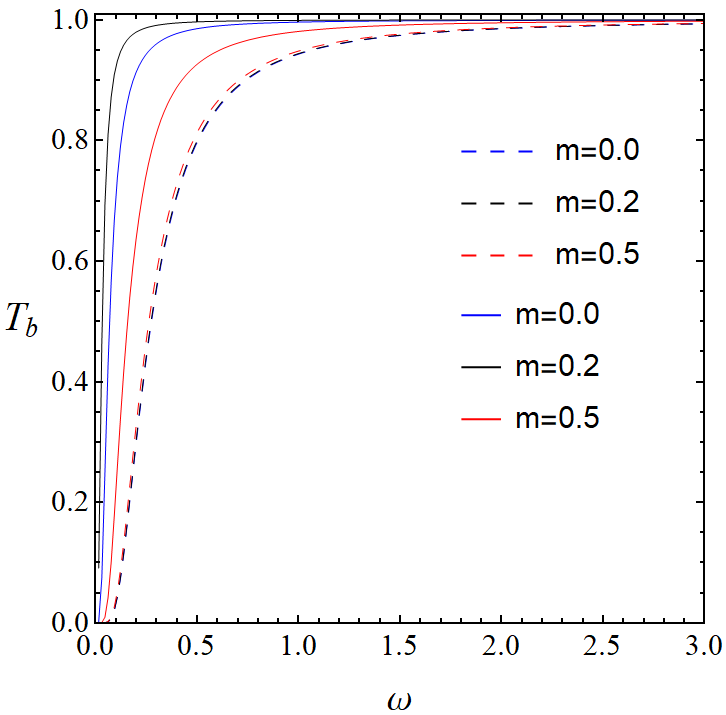}
  \caption{}
\end{subfigure}
\begin{subfigure}{.5\textwidth}
  \centering
  \includegraphics[width=0.7\linewidth]{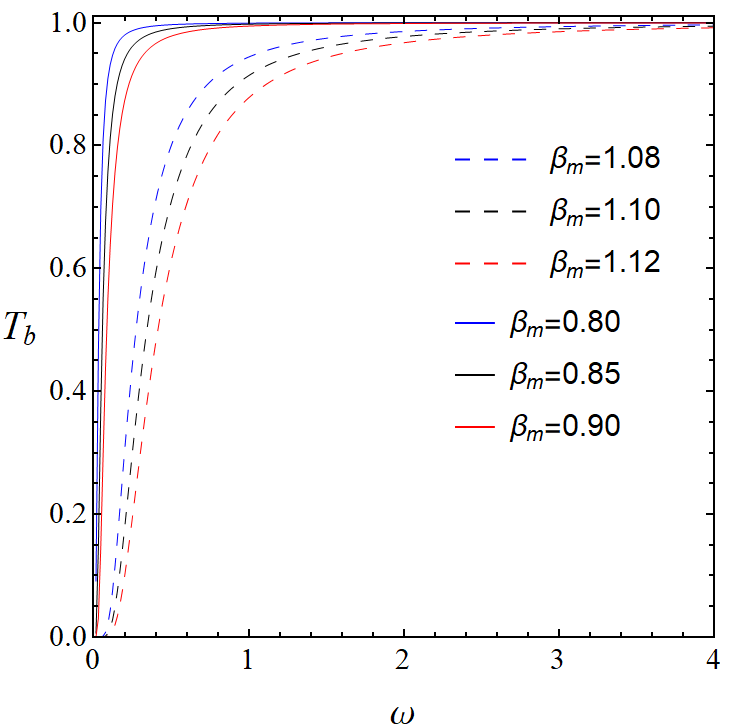}
  \caption{}
\end{subfigure}
\begin{subfigure}{.5\textwidth}
  \centering
  \includegraphics[width=0.7\linewidth]{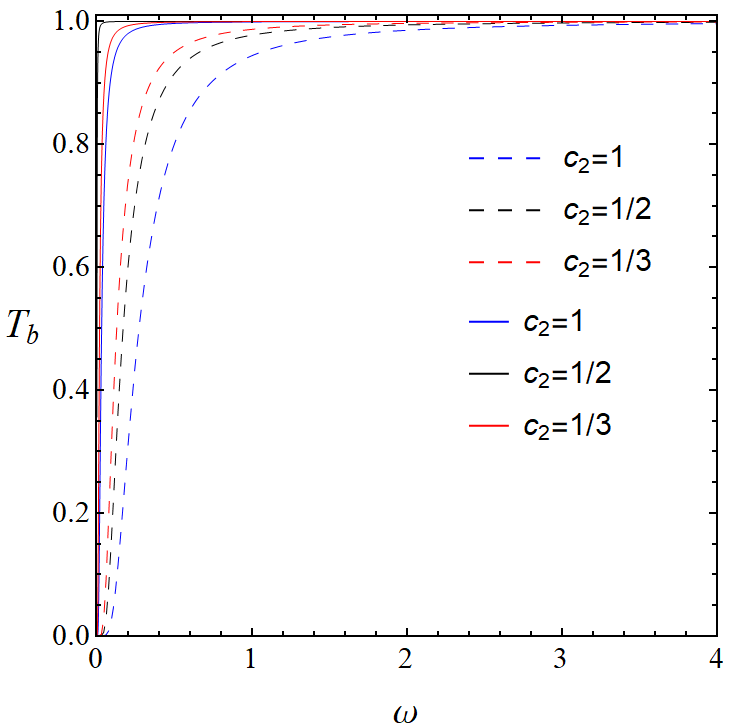}
  \caption{}
\end{subfigure}
\caption{\label{bound2}The rigorous bounds (between $\tilde{r}_{1}$ and $\tilde{r}_{2}$) on the greybody factors for the AdS case with $\ell = 0$, $\tilde{M} = 1$, $\alpha_{g} = 1$. (a) $c_{2} = 1$, $\beta_{m} = 1.08$ and $\tilde{m} = 0$, 0.2 and 0.5. (b) $c_{2} = 1$, $\tilde{m} = 0.2$ and $\beta_{m} = 1.08$, 1.10 and 1.12. (c) $\tilde{m} = 0.2$, $\beta_{m} = 1.08$ and $c_{2} = 1$, 1/2 and 1/3. The solid line represents $T_{\textrm{app}}$ and the dash line represents $T_{\textrm{b}}$.}
\end{figure}

The graphs of $T_{\textrm{app}}$ with different $\tilde{m}$, $\beta_{m}$ and $c_{2}$ are plotted in Figure \ref{bound2}(a), \ref{bound2}(b) and \ref{bound2}(c), respectively. In Figure \ref{bound2}(a), the result is interesting. When $\tilde{m}$ increases to 0.2, the bound also increases. However, if $\tilde{m}$ continues to increase, the bound starts to decrease. We see that the blue curve ($\tilde{m} = 0$) is below the grey curve ($\tilde{m} = 0.2$), but above the red curve ($\tilde{m} = 0.5$). For the low mass, the more massive scalar field can better be transmitted from the AdS potential. However, for the high mass, the more massive scalar field can be more difficult to transmit from the AdS potential.

For Figure \ref{bound2}(b), we see that when $\beta_{m}$ increases, the bound on the greybody factors decreases. This is interesting because from Figure \ref{pot4}(a), the local maximum of the potential also decreases with increasing $\beta_{m}$. The changes in the AdS potential and the AdS bound with $\beta_{m}$ are the same.

For Figure \ref{bound2}(c), the result is also interesting. We see that when $c_{2}$ increases to 1/2, the bound on the greybody factors also increases. However, if $c_{2}$ continues to increase, the bound starts to decrease. We see that the red curve ($c_{2} = 1/3$) is below the grey curve ($c_{2} = 1/2$), but above the blue curve ($c_{2} = 1$). For the background of the weak cosmological constant, the massive scalar field can better be transmitted from the AdS potential. However, for the background of the strong cosmological constant, the massive scalar field can be transmitted from the AdS potential, but with more difficulty.

Moreover, we explicitly see that $T_{\textrm{app}}$ (the solid curve) is greater than or equal to $T_{\textrm{b}}$ (the dash curve) according to inequality (\ref{TappTb}).

\begin{figure}[H]
\begin{subfigure}{.5\textwidth}
  \centering
  \includegraphics[width=0.5\linewidth]{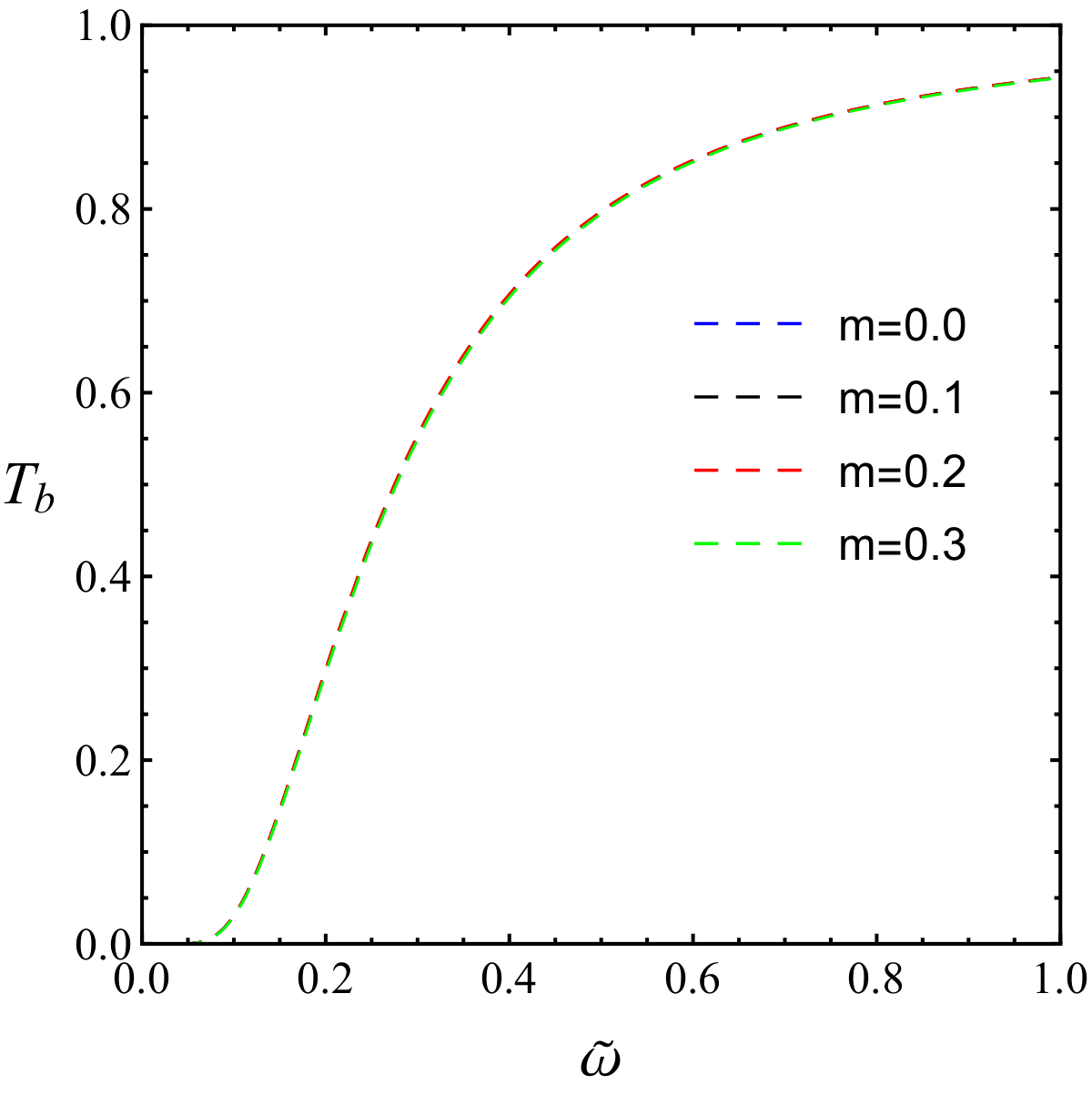}
  \caption{}
\end{subfigure}
\begin{subfigure}{.5\textwidth}
  \centering
  \includegraphics[width=0.5\linewidth]{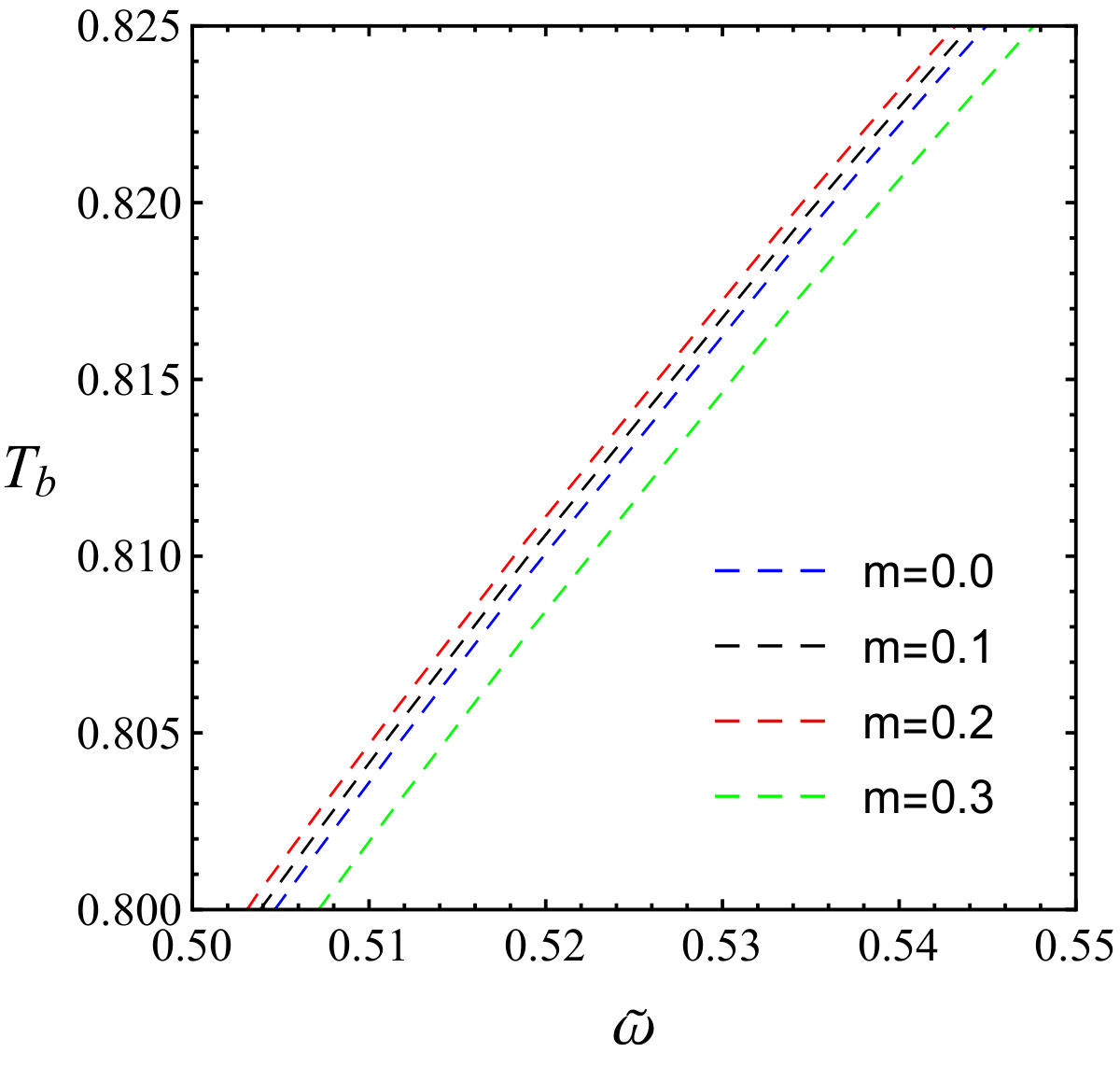}
  \caption{}
\end{subfigure}
\begin{subfigure}{.5\textwidth}
  \centering
  \includegraphics[width=0.5\linewidth]{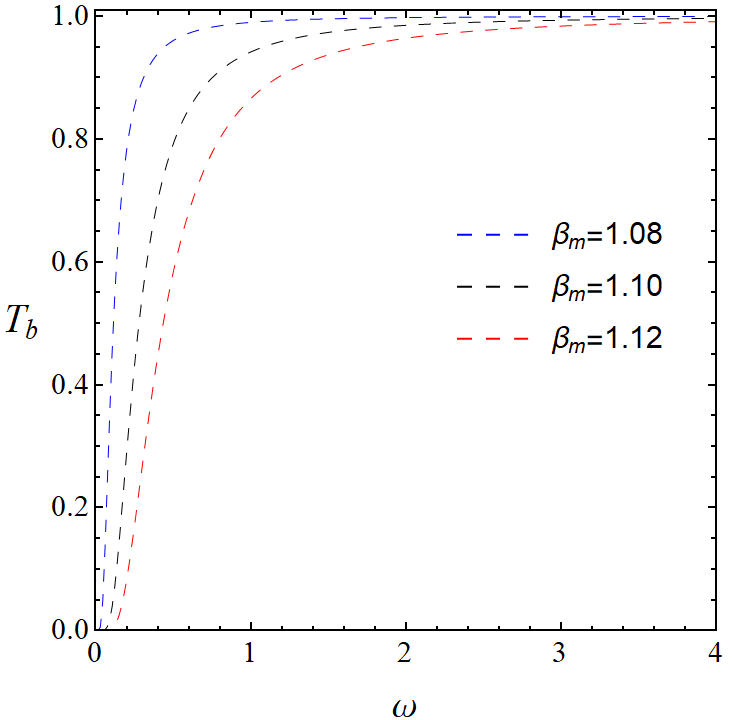}
  \caption{}
\end{subfigure}
\begin{subfigure}{.5\textwidth}
  \centering
  \includegraphics[width=0.5\linewidth]{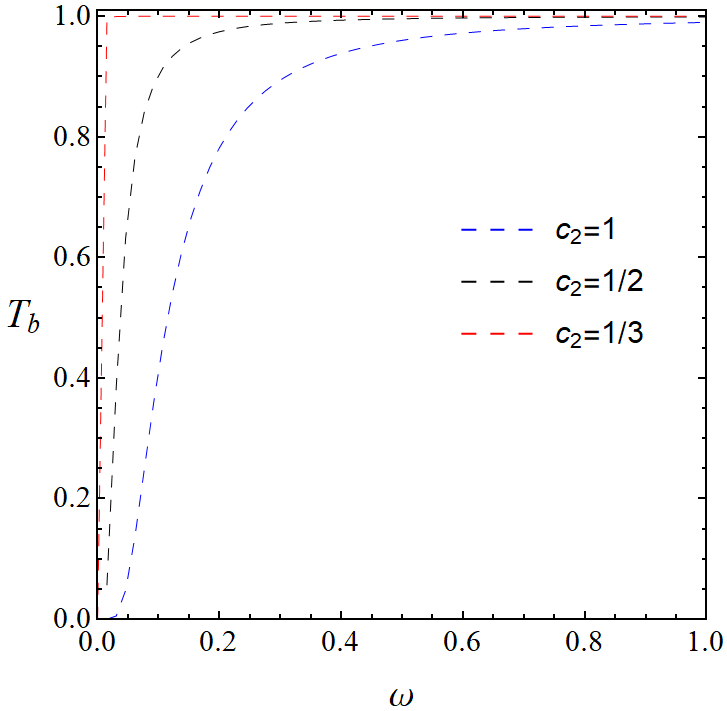}
  \caption{}
\end{subfigure}
\caption{\label{bound6}The rigorous bounds (between $\tilde{r}_{2}$ and $\tilde{r}_{3}$) on the greybody factors of the AdS black holes with $\ell = 0$, $\tilde{M} = 1$, $\alpha_{g} = 1$. (a) and (b) $c_{2} = 1$, $\beta_{m} = 1.08$ and $\tilde{m} = 0$, 0.2 and 0.5. (c) $c_{2} = 1$, $\tilde{m} = 0.2$ and $\beta_{m} = 1.08$, 1.10 and 1.12. (d) $\tilde{m} = 0.2$, $\beta_{m} = 1.08$ and $c_{2} = 1$, 1/2 and 1/3.}
\end{figure}

In the second region, where $\tilde{r}_{2} < \tilde{r} < \tilde{r}_{3}$, $f_{\textrm{AdS}}(\tilde{r})$ is negative. Moreover, the AdS potential is positive for $\tilde{r}_{2} < \tilde{r} < \tilde{r}_{\textrm{2cri}}$ and negative for $\tilde{r}_{\textrm{2cri}} < \tilde{r} < \tilde{r}_{3}$, where $\tilde{r}_{\textrm{2cri}}$ is the point satisfying $f'_{\textrm{AdS}}(\tilde{r}_{\textrm{2cri}}) = -\tilde{m}^{2}\tilde{r}_{\textrm{2cri}}$. From equation (\ref{T}), we can write
\begin{eqnarray}
T_{\textrm{AdS}} &\geq& \textrm{sech}^{2}\left[\frac{1}{2\tilde{\omega}}\int_{\tilde{r}_{2}}^{\tilde{r}_{\textrm{2cri}}}\frac{\tilde{V}_{\textrm{AdS}}(\tilde{r})}{f_{\textrm{AdS}}(\tilde{r})}d\tilde{r}\right.\nonumber\\
   && \left. - \frac{1}{2\tilde{\omega}}\int_{\tilde{r}_{\textrm{2cri}}}^{\tilde{r}_{3}}\frac{\tilde{V}_{\textrm{AdS}}(\tilde{r})}{f_{\textrm{AdS}}(\tilde{r})}d\tilde{r}\right],
\end{eqnarray}
where $\tilde{V}_{\textrm{AdS}}(\tilde{r})/f_{\textrm{AdS}}(\tilde{r})$ is
\begin{equation}
\frac{\tilde{V}_{\textrm{AdS}}(\tilde{r})}{f_{\textrm{AdS}}(\tilde{r})} = \frac{\ell\left(\ell + 1\right)}{\tilde{r}^{2}} + \frac{2\tilde{M}}{\tilde{r}^{3}} + 2\alpha_{g}c_{2} - \frac{3\alpha_{g}\sqrt[3]{4c_{2}^{2}}}{\tilde{r}} + \tilde{m}^{2}.
\end{equation}
Calculating the integrals, we obtain
\begin{eqnarray}
T_{\textrm{AdS}} &\geq& \textrm{sech}^{2}\left[\frac{1}{2\tilde{\omega}}\left\{\ell\left(\ell + 1\right)\left(\frac{1}{\tilde{r}_{2}} + \frac{1}{\tilde{r}_{3}} - \frac{2}{\tilde{r}_{\textrm{2cri}}}\right)\right.\right.\nonumber\\
      && + \tilde{M}\left(\frac{1}{\tilde{r}_{2}^{2}} + \frac{1}{\tilde{r}_{3}^{2}} - \frac{2}{\tilde{r}_{\textrm{2cri}}^{2}}\right)\nonumber\\
      && - 2\alpha_{g}c_{2}(\tilde{r}_{2} + \tilde{r}_{3} - 2\tilde{r}_{\textrm{2cri}})\nonumber\\
      && - 3\alpha_{g}\sqrt[3]{4c_{2}^{2}}\ln\frac{\tilde{r}_{\textrm{2max}}^{2}}{\tilde{r}_{2}\tilde{r}_{3}}\nonumber\\
      && \left.\left. - \tilde{m}^{2}(\tilde{r}_{2} + \tilde{r}_{3} - 2\tilde{r}_{\textrm{2cri}})\right\}\right].
\end{eqnarray}
The bounds on the greybody factors with different $\tilde{m}$, $\beta_{m}$ and $c_{2}$ are plotted in Figure \ref{bound6}(a), \ref{bound6}(c) and \ref{bound6}(d), respectively. In Figure \ref{bound6}(a) and \ref{bound6}(b), the bound increases with increasing $\tilde{m}$ until $\tilde{m}$ = 0.2. After that, the bound starts to decrease with increasing $\tilde{m}$. We see that the blue curve ($\tilde{m} = 0$) is below the grey curve ($\tilde{m} = 0.2$), but above the red curve ($\tilde{m} = 0.5$). Actually, by performing a numerical method as done in the previous case, we found the critical mass for which the greybody factor bound is maximum. For example, in our parameter setting, $c_2 = 1, \beta_m =1.08$, the critical mass can be obtained as $m_c = 0.186$. 


For Figure \ref{bound6}(c), we see that when $\beta_{m}$ increases, the bound on the greybody factors decreases. This follows from Figure \ref{pot4}(a) where the absolute of the second local maximum and the second local minimum of the potential increases with increasing $\beta_{m}$.

For Figure \ref{bound6}(d), we see that the bound on the greybody factors decreases with increasing $c_{2}$. This is the regular result because from Figure \ref{pot4}(b), the second local maximum increases with increasing $c_{2}$. In the background with the stronger cosmological constant, the massive scalar field can penetrate the AdS potential, but with more difficulty.

\begin{figure}[H]
\begin{subfigure}{.5\textwidth}
  \centering
  \includegraphics[width=0.7\linewidth]{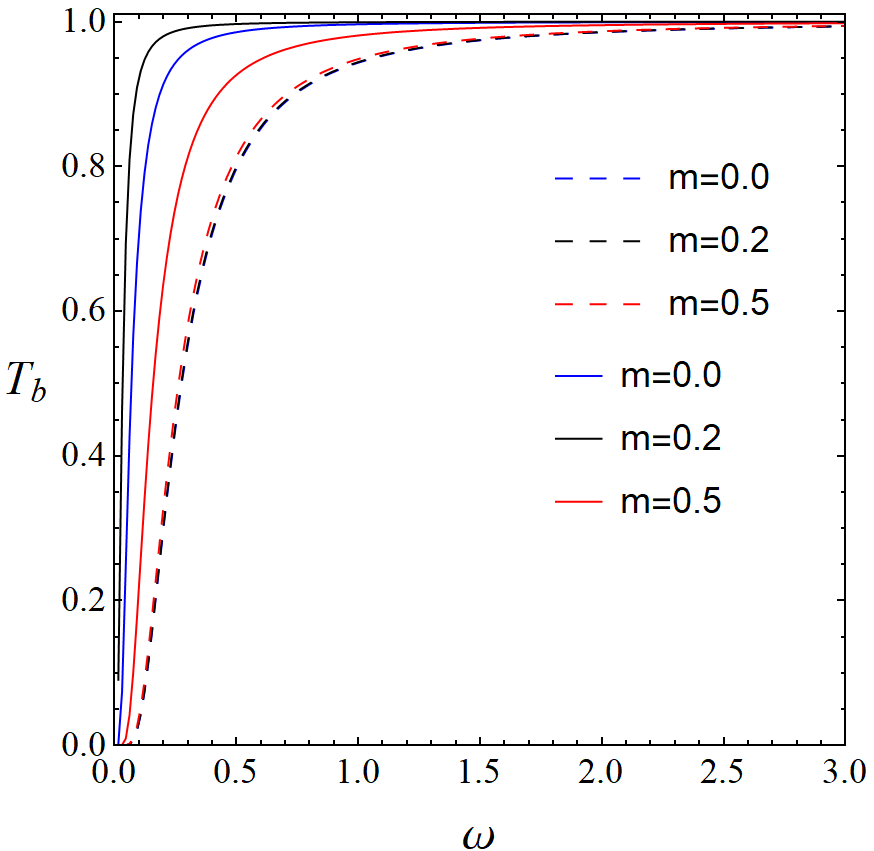}
  \caption{}
\end{subfigure}
\begin{subfigure}{.5\textwidth}
  \centering
  \includegraphics[width=0.7\linewidth]{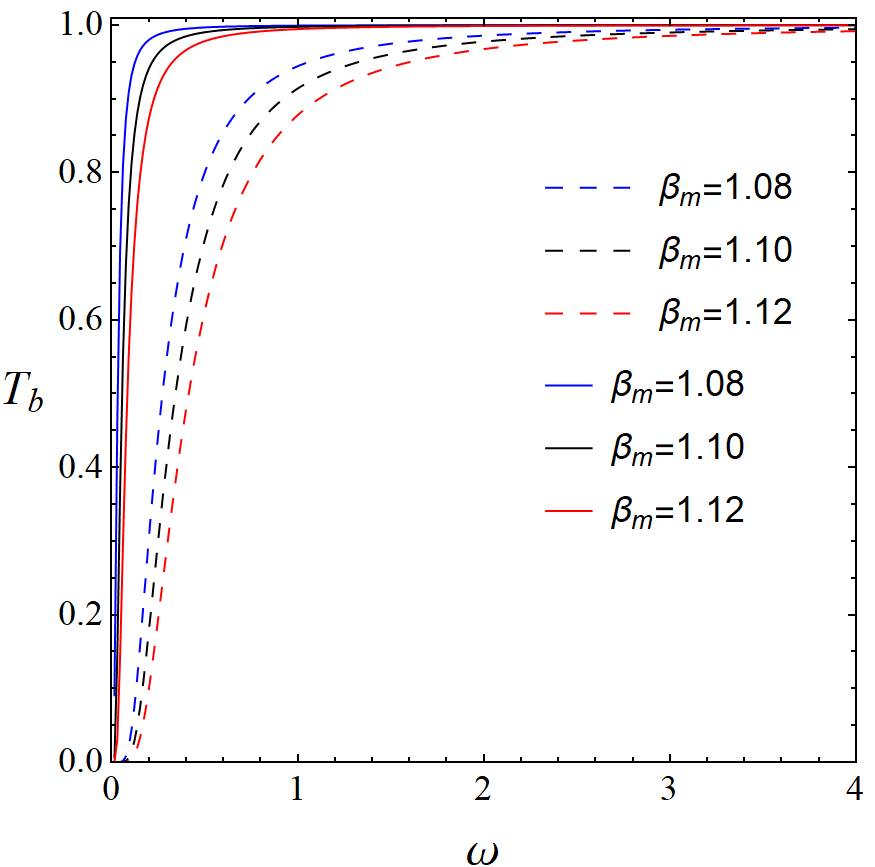}
  \caption{}
\end{subfigure}
\begin{subfigure}{.5\textwidth}
  \centering
  \includegraphics[width=0.7\linewidth]{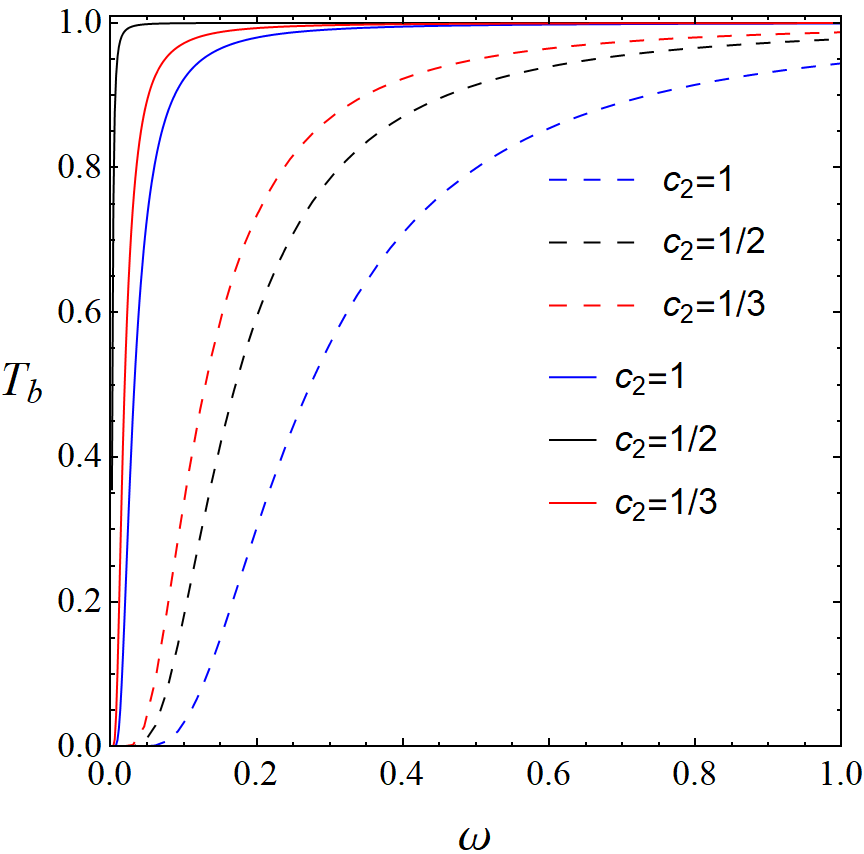}
  \caption{}
\end{subfigure}
\caption{\label{bound5}The rigorous bounds (between $\tilde{r}_{2}$ and $\tilde{r}_{3}$) on the greybody factors of the AdS black holes with $\ell = 0$, $\tilde{M} = 1$, $\alpha_{g} = 1$. (a) $c_{2} = 1$, $\beta_{m} = 1.08$ and $\tilde{m} = 0$, 0.2 and 0.5. (b) $c_{2} = 1$, $\tilde{m} = 0.2$ and $\beta_{m} = 1.08$, 1.10 and 1.12. (c) $\tilde{m} = 0.2$, $\beta_{m} = 1.08$ and $c_{2} = 1$, 1/2 and 1/3.}
\end{figure}

The graph of $T_{\textrm{app}}$ with different $\tilde{m}$, $\beta_{m}$ and $c_{2}$ are plotted in Figure \ref{bound5}(a), \ref{bound5}(b) and \ref{bound5}(c), respectively. In Figure \ref{bound5}(a), we see that the bound decreases with increasing $\tilde{m}$. This result is interesting because from Figure \ref{pot3}(a) and \ref{pot3}(b), the second local maximum also decreases with increasing $\tilde{m}$. The changes in the second local maximum of the AdS potential and the AdS bound with $\tilde{m}$ are the same. The more massive scalar field can be harder to transmit from the AdS potential.

For Figure \ref{bound5}(b), we see that when $\beta_{m}$ increases, the bound on the greybody factors also increases. This is interesting because from Figure \ref{pot4}(a), the second local maximum of the potential also increases with increasing $\beta_{m}$. Again, the changes in the second local maximum of the AdS potential and the AdS bound with $\beta_{m}$ are the same.

For Figure \ref{bound5}(c), we see that the bound on the greybody factors decreases with increasing $c_{2}$. This is the regular result because from Figure \ref{pot4}(b), the second local maximum increases with increasing $c_{2}$. In the background with the stronger cosmological constant, the massive scalar field can penetrate the AdS potential, but with more difficulty.

Moreover, we explicitly see that $T_{\textrm{app}}$ (the solid curve) is greater than or equal to $T_{\textrm{b}}$ (the dash curve) according to the inequality (\ref{TappTb}).

\subsubsection{$\ell \neq 0$}
From Figure \ref{potadsl12}, the AdS potential for $\ell \neq 0$ is positive on the interval $\tilde{r}_{1} < \tilde{r} < \tilde{r}_{2}$ and negative on the interval $\tilde{r}_{2} < \tilde{r} < \tilde{r}_{3}$. That is the AdS potential for $\ell \neq 0$ has the same sign as $f_{\textrm{AdS}}(\tilde{r})$. Therefore, from equations (\ref{Tapp}) and (\ref{Tb}), $T_{\textrm{b}} = T_{\textrm{app}}$. Then, equation (\ref{TTb}) becomes
\begin{equation}
T_{\textrm{AdS}, \ell \neq 0} \geq T_{\textrm{b}, \textrm{AdS}, \ell \neq 0},
\end{equation}
where
\begin{equation}
T_{\textrm{b}, \textrm{AdS}, \ell \neq 0} = \textrm{sech}^{2}\left[\frac{1}{2\tilde{\omega}}\int_{\tilde{r}_{1}}^{\tilde{r}_{2}}\frac{\tilde{V}_{\textrm{AdS}, \ell \neq 0}(\tilde{r})}{f_{\textrm{AdS}}(\tilde{r})}d\tilde{r}\right]
\end{equation}
and
\begin{equation}
\frac{\tilde{V}_{\textrm{AdS}, \ell \neq 0}(\tilde{r})}{f_{\textrm{AdS}}(\tilde{r})} = \frac{\ell(\ell + 1)}{\tilde{r}^{2}} + \frac{2\tilde{M}}{\tilde{r}^{3}} + 2\alpha_{g}c_{2} - \frac{3\alpha_{g}\sqrt[3]{4c_{2}^{2}}}{\tilde{r}} + \tilde{m}^{2}.
\end{equation}
Calculating the integrals, we obtain
\begin{eqnarray}
T_{\textrm{b}, \textrm{AdS}, \ell \neq 0} &=& \textrm{sech}^{2}\left[\frac{1}{2\tilde{\omega}}\left\{\ell(\ell + 1)\left(\frac{1}{\tilde{r}_{1}} - \frac{1}{\tilde{r}_{2}}\right)\right.\right.\nonumber\\
      && + \tilde{M}\left(\frac{1}{\tilde{r}_{1}^{2}} - \frac{1}{\tilde{r}_{2}^{2}}\right) + 2\alpha_{g}c_{2}(\tilde{r}_{2} - \tilde{r}_{1})\nonumber\\
      && \left.\left. - 3\alpha_{g}\sqrt[3]{4c_{2}^{2}}\ln\frac{\tilde{r}_{2}}{\tilde{r}_{1}} + \tilde{m}^{2}(\tilde{r}_{2} - \tilde{r}_{1})\right\}\right].
\end{eqnarray}

\begin{figure}[H]
\begin{subfigure}{.5\textwidth}
  \centering
\includegraphics[width=0.7\linewidth]{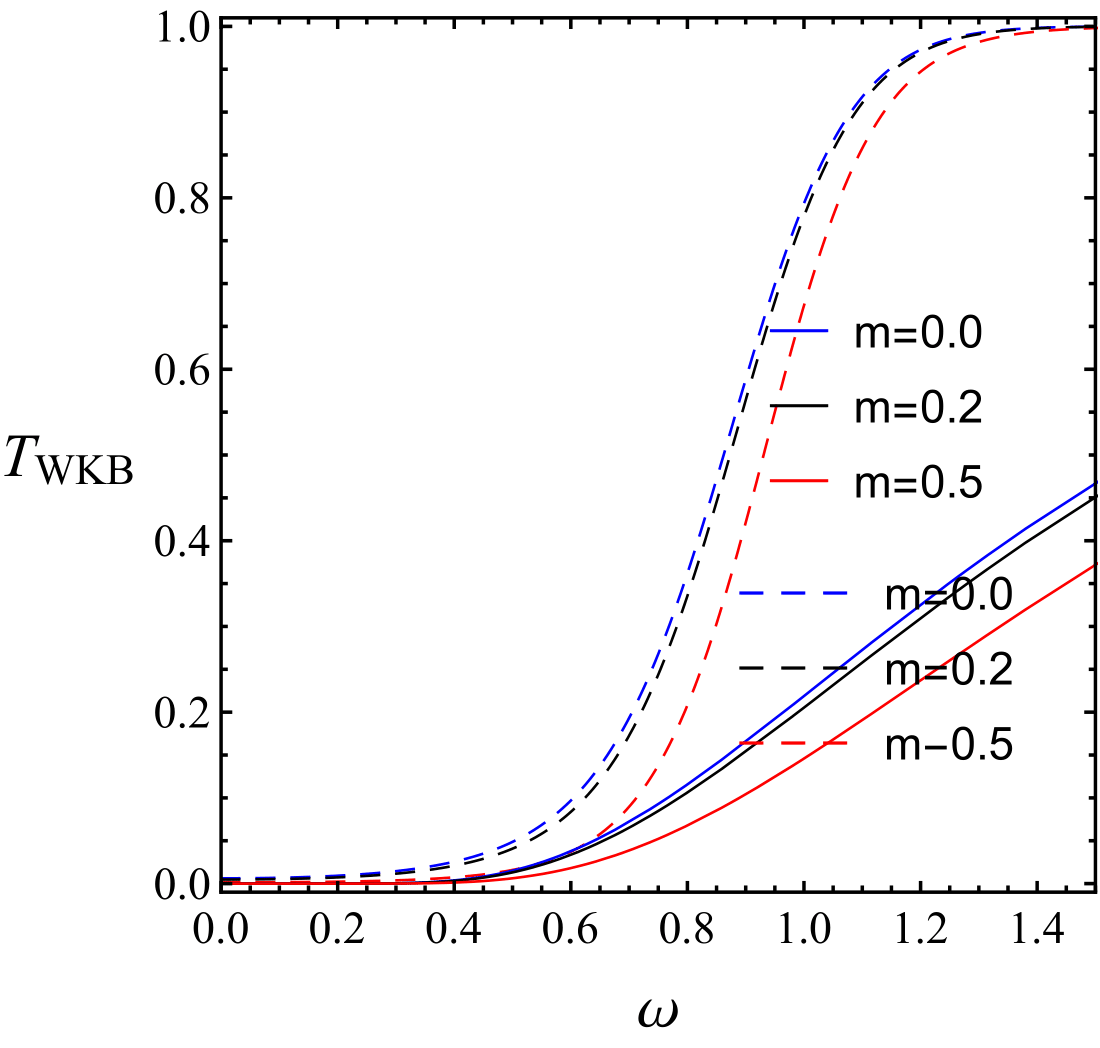}
  \caption{}
\end{subfigure}
\begin{subfigure}{.5\textwidth}
  \centering
 \includegraphics[width=0.7\linewidth]{VAdS_m_L2.png}
  \caption{}
\end{subfigure}
\caption{\label{VdGRTl1-AdS2}(a) The greybody factor between $\tilde{r}_{2}$ and $\tilde{r}_{3}$ of the AdS black holes, $T_{\textrm{b}}$ from the bound (solid line) and $T_{\textrm{WKB}}$ from the WKB method (dashed line) for $c_{2} = 1, \beta_{m} = 1.1$, $\ell = 2$ and $\tilde{m} = 0$, 0.2 and 0.5. (b) The AdS potential with $c_{2} = 1, \beta_{m} = 1.1$ and $\ell = 2$.}
\end{figure}

The graphs of $T_{\textrm{WKB}}$ compared to $T_{\textrm{b}}$ between $\tilde{r}_{2}$ and $\tilde{r}_{3}$ with different $\tilde{m}$ are plotted in Figure \ref{VdGRTl1-AdS2}. Note that the results for one  between $\tilde{r}_{1}$ and $\tilde{r}_{2}$ are  similar to this case and we omit to present here. From this Figure, we see that $T_{\textrm{b}}$ is less than $T_{\textrm{WKB}}$, showing that $T_{\textrm{b}}$ is a true lower bound. Note that they are so close to one another when $\tilde{\omega}$ is around 0.4. Moreover, their behaviors are similar. When $\tilde{m}$ increases, both $T_{\textrm{WKB}}$ and $T_{\textrm{b}}$ decrease.



\section{Conclusion}\label{con}
One of the interesting issues of a black hole is that it behaves as a thermal system. Specifically, a black hole can carry entropy and can emit radiation called the Hawking radiation. Moreover, the spectrum of the radiations from black holes is the same as that of the black-body spectrum. It has also been found that the spacetime curvature can act as a potential barrier, which allows some of the radiation to transmit and reflect as found in similar situations in quantum mechanics. As a result, the greybody factor is defined in order to take into account the transmission amplitude of the radiation from the black holes.

In this paper, we investigate the greybody factor for the massive scalar field from the dRGT black hole. In theories beyond GR, there are usually additional degrees of freedom that can be coupled to gravity. A simplest example is known as the scalar-tensor 
theory, in which a scalar field has couplings to gravity. In this sense, the scalar field may be one of the simple candidates for the Hawking radiation, making it an interesting issue to investigate the greybody factor for a massive scalar field. One of the interesting modified gravity theories is the dRGT massive gravity, since it is a candidate for describing the late-time expansion of the universe. Moreover, it turns out that there exists a black hole solution in the dRGT massive gravity. One of the interesting properties of the dRGT black hole is that it is possible to obtain the asymptotically dS/AdS spacetime, which can be distinguished from the SdS/SAdS black hole. Therefore, it is worthwhile to investigate the greybody factor from the dRGT black hole by analyzing how the mass of the scalar field influences the greybody factor. 

In the present work, we begin with the review of the dRGT black hole, including the horizon structure of the black hole in both the asymptotically dS and the AdS spacetime. For the dRGT black hole with the asymptotically dS spacetime, there are generically two horizons. In order to characterize the existence of the horizon, we introduce the model parameter $\beta_{m}$ in which $0 < \beta_{m} < 1$ for the existence of two horizons. This parameter also influences the structure of the horizon as the larger the value of the parameter, the narrower the width of the two horizons. Moreover, the strength of the cosmological constant can be characterized by parameter $c_{2}$. For the dS case, when $c_{2} < 0$ the role of $c_{2}$ is the same as the cosmological constant. For example, the smaller the value of $|c_{2}|$, the wider width of the horizons. For the AdS case, we focus on the black hole with three horizons, which is distinguished from the SAdS black hole (only one horizon exists). In the same manner as the dS case, we are left with only two parameters to characterize the horizon structure where $1 < \beta_{m} < 2/\sqrt{3}$ is the condition for the existence of three horizons.

We find the potential, which acts as the barrier allowing the scalar field to transmit in a similar manner to quantum mechanics. Since the potential depends on the horizon function and its derivative, it is controlled by two parameters $\beta_{m}$ and $c_{2}$. Moreover, there are two more parameters, the scalar field mass $m$ and the multipole $\ell$, to characterize the behaviour of the potential. It is found that the higher the value of the scalar field mass, the higher the value of the maximum potential. In fact, for both the dS and the AdS cases, the mass term contributes to the positive part of the potential, to shift the potential upward. For the dS case, it is found that the potential from the dRGT black hole is narrower than one from the SdS black hole. This behaviour is inherited from the narrow width of the two horizons. For the AdS case, the potential from the dRGT black hole cannot be compared to one from SAdS, since the structure of the horizons are completely different. As for the multipole term, it also provides a positive contribution to the potential for which the potential will be shifted upward where the parameter $\ell$ is increased. It is found that for $\ell=0$ with a small scalar field mass, there exists a negative part of the potential. In this case, we have to carefully consider the calculation for the greybody factor bound since the expression contains the absolute value of the potential.  

The greybody factors from various kinds of spacetime geometry have been intensively investigated by various methods. One of the ways is finding solutions in the asymptotic regions and then matching the solutions at the boundaries. However, the solutions are mostly written in terms of special functions, which makes it difficult to analyze the behaviour of the spectrum analytically. Other interesting way, which is intensively investigated in literature, is that of using the WKB approximation. It provides a good approximation for a simple form of spacetime geometries, which then requires the higher potential, or in other words, requires high multipole. The other way to investigate the greybody factor is to consider the bound of the greybody factor instead of the exact one. This method allows us to study the behaviour of the greybody factor analytically.

For the dS case, the greybody factor depends on the shape of the potential as found in quantum mechanics. The scalar field mass provides a higher potential so that it is more difficult to transmit. Therefore, this provides a lower greybody factor. However, for the case $\ell = 0$ with a small scalar field mass, there exists a negative part of the potential. We found that there exists a critical mass which provides the maximum bound of the greybody factor. This is one of the crucial behaviour of the massive scalar field for the greybody factor. Note that this can be found in both the dRGT black hole and the SdS black hole. Moreover, we found that the greybody factor from the dRGT black hole is higher than one from the SdS black hole. This is due to the fact that the potential for the dRGT black hole is thinner than one for the SdS black hole. It is also found that the results of the greybody factor bound agree with ones from the WKB method.

For AdS case, the role  of the scalar field mass is similar to the dS case in both systems between $r_1$ and $r_2$, and $r_2$ and $r_3$. The higher the mass, the higher the potential and the lower the greybody factor. However, for the case in which there exist both the negative and the positive part of the potential, e.g. $\ell = 0$ with the small scalar field mass, we found that there exists a critical mass which provides the maximum bound of the greybody factor. This is one of the crucial behaviour of the massive scalar field for the greybody factor. We also found that the results of the greybody factor bound agree with ones from the WKB method.

\medskip

{\bf Acknowledgment.} This research has received funding support from the NSRF via the Program Management Unit for Human Resources \& Institutional Development, Research and Innovation [grant number B37G 660013]. This project was funded by the Ratchadapisek Somphot Fund, Chulalongkorn University, for the professional development of new academic staff, by Thailand NSRF via PMUB [grant number B05F650021], by the Thailand Research Fund (TRF), by the Office of the Higher Education Commission (OHEC), and supported by the Fundamental Fund from the National Science, Research and Innovation Fund (NSRF) under the grant ID RSA2565B030, Faculty of Science, Chulalongkorn University (RSA5980038). PB was additionally supported by a scholarship from the Royal Government of Thailand. SP was also additionally supported by a scholarship from the Development and Promotion of Science and Technology Talents Project (DPST). KS was also additionally supported by the scholarship committee from the Thailand Center of Excellence in Physics at the Ministry of Higher Education in Science, Research and Innovation.

\end{document}